\documentclass{article}

\usepackage{arxiv}

\usepackage[utf8]{inputenc} 
\usepackage[T1]{fontenc}    
\usepackage[hidelinks]{hyperref}       
\usepackage{url}            
\usepackage{booktabs}       
\usepackage{amsmath}
\usepackage{amsfonts}       
\usepackage{nicefrac}       
\usepackage{microtype}      
\usepackage{lipsum}
\usepackage{graphicx}
\graphicspath{ {./images/} }
\usepackage{subfig}
\usepackage{multirow}
\usepackage{natbib}

\title{Incorporating Magnetic Field Characteristics into EUV-Based Automated Segmentation of Coronal Holes}

\author{Jeremy A.~Grajeda, Laura E.~Boucheron \\
  Klipsch School of Electrical and Computer Engineering\\
  New Mexico State University\\
  Las Cruces, NM 88003, USA \\
  \texttt{\{jgra,lboucher\}@nmsu.edu} \\
   \And
  Michael S.~Kirk \\
  Heliophysics Space Division, Goddard Space Flight Center \\
  National Aeronautics and Space Administration \\
  Greenbelt, MD 20771, USA\\
  \texttt{michael.s.kirk@nasa.gov} \\
  \And
  Andrew Leisner \\
  Department of Physics and Astronomy \\
  George Mason University \\
  Fairfax, VA 22030, USA \\
  \texttt{aleisner@gmu.edu} \\
  \And
  C.~Nick Arge \\
  Heliophysics Space Division, Goddard Space Flight Center \\
  National Aeronautics and Space Administration \\
  Greenbelt, MD 20771, USA\\
  \texttt{charles.n.arge@nasa.gov} \\
  \And
  Jaime~A.~Landeros \\
  Heliophysics Space Division, Goddard Space Flight Center \\
  National Aeronautics and Space Administration \\
  Greenbelt, MD 20771, USA\\
  University of California San Diego\\
  La Jolla, CA 92093, USA \\
  ADNET Systems Inc\\
  Bethesda, MD 20817, USA\\
  \texttt{jalanderos@ucsd.edu} \\
}

\begin{document}
\maketitle\let\thefootnote\relax\footnotetext{This version of the article has been accepted for publication, after peer review, but is not the Version of Record and does not reflect post-acceptance improvements, or any corrections.  The Version of Record is available online at \url{https://doi.org/10.1007/s11207-025-02536-7}}
\begin{abstract}
Coronal holes (CHs) are magnetically open regions that allow hot coronal plasma to escape from the Sun and form the high-speed solar wind. This wind can interact with Earth’s magnetic field. For this reason, developing an accurate understanding of CH regions is vital for understanding space weather and its effects on Earth. The process of identifying CH regions typically relies on extreme ultraviolet (EUV) imagery, leveraging the fact that CHs appear dark at these wavelengths. Accurate identification of CHs in EUV, however, can be difficult due to a variety of factors, including stray light from nearby regions, limb brightening, and the presence of filaments (which also appear dark, but are not sources of solar wind). In order to overcome these issues, this work incorporates photospheric magnetic field data into a classical EUV-based segmentation algorithm based on the active contours without edges (ACWE) segmentation method. In this work magnetic field data are incorporated directly into the segmentation process, serving both as a method for removing non-CH regions in advance, and as a method to constrain evolution of the segmented CH boundary. This reduces the presence of filaments while allowing the segmentation to include CH regions that may be difficult to identify due to inconsistent intensities.
\end{abstract}


\section{Introduction}
Coronal holes (CHs) are regions of open magnetic field lines \citep{Altschuler1972}. These lines, which emerge from the photosphere and extend through the Sun's corona, allow coronal plasma to escape from the Sun and form high speed flows of particles called the solar wind \citep{Wang1990,Wang1996,Antonucci2004,McComas2007} that can affect Earth's magnetic field \citep{tsurutani2006}. Due to this relationship between CHs and solar wind, accurate detection and segmentation of CHs proves vital for understanding solar wind behavior, and, by extension, improving space weather forecasting. This can be seen, for example, in the development of the Wang-Sheeley-Arge (WSA) model \citep{Wang1990,Arge2003,Arge2004}.

Due to the lower density and temperature of CH regions, a byproduct of the open magnetic field lines, CHs can be detected in extreme ultraviolet (EUV) imagery, where they appear as dark regions \citep{Munro1972}. This has resulted in a large number of detection schemes which rely on EUV imagery, including those developed by \cite{Krista2009, verbeeck2014, Lowder2014, Caplan2016, boucheron2016segmentation, Garton2018, Hamada2018, grajeda2023}. Each of these methods addresses difficulties with the segmentation process, including inconsistent intensities caused by stray light from nearby regions, limb brightening, and instrument effects \citep{verbeeck2014, Caplan2016}. In addition to these difficulties, the presence of filaments, which also appear dark in EUV but are not sources of solar wind, can result in the misclassification of regions \citep{Krista2009, reiss2023}. 

The open magnetic field lines present in CHs emerge from highly unipolar regions unlike magnetic fields in filaments which are bipolar in nature. This has resulted in several methods that validate CH regions after detection through the use of magnetic field data, removing non-CH regions as a post-hoc process \citep{Scholl2008, Krista2009, Lowder2014, Garton2018, Hamada2018, landeros2024}. Direct application of magnetic field data in the initial segmentation of CHs remains limited, appearing predominantly in \cite{Jarolim2021} where it serves as an additional channel alongside seven EUV channels input to a network based on the U-Net architecture of \cite{Ronneberger2015}. The work presented here proposes incorporating magnetic field data into the classical segmentation method first employed by \cite{boucheron2016segmentation} and extended in \cite{grajeda2023}, which is based on Active Contours Without Edges (ACWE) \citep{chan2001}. This ACWE-based method has been shown to effectively overcome inconsistent intensities present in CH observations for the majority of cases and is robust across short timeframes where CH evolution is expected to be minimal \citep{grajeda2023}. However, like all other EUV-based methods, it can misidentify filament regions \citep{grajeda2023,reiss2023}. Through the addition of magnetic field data in both seeding and iterating of the ACWE process, the method proposed here reduces filament contamination, first by eliminating many filaments outright and second by reducing the area of filaments that cannot be outright removed.  Additionally, incorporation of magnetic field better constrains the evolution to avoid including too much quiet Sun (QS) as CH.

This paper is organized as follows: Section~\ref{sec:background} provides an introduction to ACWE segmentation, outlines how it was adapted for CH detection, and introduces the datasets. Section~\ref{sec:QUACK_evolve} outlines the incorporation of magnetic field data into ACWE evolution and explores the resulting effects. Section~\ref{sec:Seed} outlines the incorporation of magnetic field data into the initial seeding to help remove filament contamination outright. The final pipeline is discussed and validated in Section~\ref{sec:pipelineSpecific}. Conclusions and discussion of future work are provided in Section~\ref{sec:conclusion}. 

\section{Background}
\label{sec:background}

\subsection{Active Contours Without Edges (ACWE)}

The detection method proposed here is an extension of the ACWE algorithm developed by \cite{chan2001} and first applied to CH detection in \cite{boucheron2016segmentation}. ACWE is a two-step process that separates an image into foreground (CHs in our case) and background (remaining solar features and off-disk areas). In the first step, one or more enclosed shapes, collectively called the contour ($C$), are defined. This contour acts as an initial guess, with regions inside the contour ($C^+$) serving as the foreground and regions outside the contour ($C^-$) as the background. In the second step, the contour is refined by evaluating every pixel along the contour boundary to determine if it is more similar to the pixels within $C^+$ or $C^-$ and redrawing $C$ accordingly. The second step is repeated until a stopping criterion is met.

In the standard formulation of ACWE, determining whether a pixel along the contour boundary belongs to $C^+$ or $C^-$ is decided using the energy functional:
\begin{equation}
 \begin{split}
   F(m^+,m^-,C)=\mu \ell(C)+\lambda^{\rm{H}+}\int_{C^+}|I(x,y)-m^+|^2\mathrm{d}x\mathrm{d}y\\
   +\lambda^{\rm{H}-}\int_{C^-}|I(x,y)-m^-|^2\mathrm{d}x\mathrm{d}y,
 \end{split}
 \label{eq:acwe}
\end{equation}
where $I$ is the image. This equation summarizes three ``forces'' that act on $C$, giving each force a user-defined weight that determines its relative importance. The contour is manipulated to minimize the functional. The first force,
\begin{equation}
    \mu \ell(C),
    \label{eq:acwe_len}
\end{equation}
attempts to minimize the length of the contour $\ell(C)$, and has user-defined weight $\mu$ to determine its relative importance. The second force,
\begin{equation}
    \lambda^{\rm{H}+}\int_{C^+}|I(x,y)-m^+|^2\mathrm{d}x\mathrm{d}y,
    \label{eq:acwe_inside}
\end{equation}
attempts to create a ``homogeneous'' foreground, where homogeneous is understood to mean a narrow range of intensities. This is done by comparing every pixel of $I(x,y)$ in $C^+$, to the mean intensity of the foreground $m^+$. This force is subject to user-defined weight $\lambda^{\rm{H}+}$. The final force,
\begin{equation}
    \lambda^{\rm{H}-}\int_{C^-}|I(x,y)-m^-|^2\mathrm{d}x\mathrm{d}y,
    \label{eq:acwe_outside}
\end{equation}
mirrors the behavior of the second force (Equation~\ref{eq:acwe_inside}), this time trying to create a homogeneous background by comparing every pixel in $C^-$ to the mean intensity of the background $m^-$. It is subject to user-defined weight $\lambda^{\rm{H}-}$. By varying the relationship between the two homogeneity parameters $\lambda^{\rm{H}+}$ and $\lambda^{\rm{H}-}$, the user can prioritize enforcing homogeneity in either $C^+$ or $C^-$. It should be noted that the foreground, which consists of CHs, is expected to be more homogeneous than the aggregate of all remaining solar features (including QS, filaments, and active regions). For this reason, both \cite{boucheron2016segmentation} and \cite{grajeda2023} enforce homogeneity in $C^+$, and allow for inhomogeneity in $C^-$, by co-defining $\lambda^{\rm{H}+}$ and $\lambda^{\rm{H}-}$ as a ratio $\lambda^{\rm{H}+}/\lambda^{\rm{H}-}\geq10$.

In both \cite{boucheron2016segmentation} and \cite{grajeda2023}, the initial contour (seed) is defined using an intensity threshold applied to a 193~{\AA} EUV image that has been corrected for limb brightening using the method from \cite{verbeeck2014}. This threshold defines all on-disk pixels with an intensity $\leq\alpha\times m_{\rm QS}$ (where $\alpha$ is a user-defined parameter [typically $0.3$] and $m_{\rm QS}$ is an estimate of the mean intensity of the QS) as CHs. The contour is evolved using the same image, after masking off-disk areas from the algorithm. The process is stopped if no evolution occurs between iterations or if the only evolution that occurs consists of pixels along the boundary that oscillate between $C^+$ and $C^-$. 

\subsection{Dataset and Metrics}
\subsubsection{Primary Dataset and Metrics}

For this work we utilize the dataset we previously developed in \cite{grajeda2023}. This dataset contains AIA Level 1 EUV images at 94, 131, 171, 193, 211, 304, and 335~{\AA}, and the temporally corresponding 720-second line-of-sight (LOS) HMI magnetograms, collected at a one-hour cadence from Carrington rotations (CRs) 2099, 2100, 2101, and 2133. All magnetogram data have an observation time within $\pm2$~s of the record time for the corresponding EUV data. We note that LOS magnetograms may report inaccurate magnetic field at the solar limb, but our method should be able to equivalently use radial magnetic field from vector magnetograms (an area of future research). Our workflow will be the reverse of the process in \cite{grajeda2023}, tuning our proposed variant of ACWE on CR 2133 which has a high level of filament activity, to ensure that this algorithm can minimize false positive detections. Then, observations from CRs 2099-2101 will be used to verify that the algorithm continues to perform effectively during periods of low solar activity where it previously excelled. 

To accurately compare between \cite{grajeda2023} and this work, we utilize the same four metrics: intersection over union (IOU), structural similarity index measure (SSIM), global consistency error (GCE), and local consistency error (LCE). Both IOU and SSIM report a value of 1 when two segmentations are identical, with IOU serving as a stringent metric quantifying overlap between the two segmentations normalized by total area and SSIM providing a perceptual measure more consistent with human vision \citep{wang2004image}. Both GCE and LCE report a value of 0 when two segmentations are identical. GCE and LCE report minimal or no error when one segmentation is a refinement of the other, with GCE being a more stringent metric \citep{martin2001}.

\subsubsection{Secondary Datasets and Metrics}

In addition to the dataMSEE and MSEE and set of \cite{grajeda2023}, used to develop this method, two other datasets will be used in validation. The first dataset is from \cite{reiss2023}. This dataset provides labels for both CHs and filaments over 29 observations from July 2014 through June 2019. This allows for a direct comparison between the previous formulation of ACWE \citep{boucheron2016segmentation, grajeda2023} and the formulation presented here. For this comparison we provide the total number of CHs, filaments, and other objects detected by each method. We also provide a comparison of the total area, in pixels and Mm$^2$, for each filament still present in the updated formulation.

Second, we have collected a dataset with daily cadence that contains observations from the start of CR 2099 through the end of CR 2294 (13 July 2010 through 2 March 2025). This dataset is used to verify behavior of the proposed method by measuring the total area in Mm$^2$ and magnetic field underlying each segmented CH.  This determines if the behavior of objects detected by our proposed method is consistent with expectations over the portions of solar cycles 24 and 25 that SDO has captured. For this dataset we collect the seven AIA observations with record time closest to 12:00:00 on each day for which all observations have \verb+QUALITY+ key of \verb+0+ and there exists a 720-second LOS HMI magnetogram with \verb+QUALITY+ key of \verb+0+ and observation time within $\pm2$~s of the AIA record time. We also collect that 720-second LOS HMI magnetogram.

\section{Quantifying Unipolarity via Active Contour Kinetics}
\label{sec:QUACK_evolve}

Incorporation of magnetic field data into ACWE evolution requires two adjustments to the standard formulation of ACWE. First, the ACWE algorithm must be altered to allow for vector-valued images so that EUV and magnetogram data can be simultaneously considered. Second, the ACWE energy functional must be altered to maximize unipolarity as expressed by the magnetic field data while still maximizing homogeneity in the EUV observation(s).

\subsection{ACWE for Vector-Valued Images}

The vector-valued formulation of ACWE was developed by \cite{chan2000130} by redefining the ACWE energy functional as
\begin{equation}
 \begin{split}
  F(\Bar{m}^+,\Bar{m}^-,C)=\mu \ell(C)+\int_{C^+}\frac{1}{N}\sum_{i=1}^{N}\lambda^{\rm{H}+}_i|I_i(x,y)-m^+_i|^2\mathrm{d}x\mathrm{d}y\\
  +\int_{C^-}\frac{1}{N}\sum_{i=1}^{N}\lambda^{\rm{H}-}_i|I_i(x,y)-m^-_i|^2\mathrm{d}x\mathrm{d}y.
 \end{split}
 \label{eq:acwev}
\end{equation}
This new formulation evaluates the homogeneity of $C^+$ and $C^-$ on a per-channel basis by comparing the pixels (foreground and background, respectively) in each channel $I_i$ of the vector-valued image $I$ to the channel-specific means $m^+_i$ and $m^-_i$. Within this formulation the weights for the two homogeneity parameters are also defined on a per-channel basis as $\lambda^{\rm{H}+}_i$ and $\lambda^{\rm{H}-}_i$ (foreground and background, respectively), allowing the user to adjust the relative influence of each channel's foreground and background. It should be noted that when the total number of channels $N=1$, Equation~\ref{eq:acwev} reduces to Equation~\ref{eq:acwe}.

\subsection{Incorporating Magnetic Unipolarity}

The vector-valued version of ACWE only considers homogeneity within each channel. For this reason Equation~\ref{eq:acwev} must be modified to consider homogeneity in EUV only, while considering unipolarity in the corresponding magnetograms. Forcing ACWE to ignore homogeneity when evaluating the magnetic field can be accomplished by setting $\lambda^{\rm{H}+}_j=0$ and $\lambda^{\rm{H}-}_j=0$ where $j$ is the index of the magnetogram data within the vector-valued image $I$. From there, a metric that minimizes when a region becomes more unipolar can be added to Equation~\ref{eq:acwev}.

We use the unipolarity metric of \cite{Ko_2014}:
\begin{equation}
    U=\frac{\langle|B|\rangle-|\langle B\rangle|}{\langle|B|\rangle},
    \label{eq:uni}
\end{equation}
where $\langle\cdot\rangle$ denotes mean, $|\cdot|$ denotes absolute value, and $B$ is the magnetic field. This metric is bound to the range $[0,1]$, where $0$ indicates a purely unipolar region and $1$ indicates a purely bipolar region. This ensures that increasing unipolarity will minimize the energy functional.

In order to incorporate Equation~\ref{eq:uni} into Equation~\ref{eq:acwev}, the effect of moving a pixel at the contour boundary into or out of the foreground (CH) must be quantified. This means that both $\langle|B|\rangle$ and $|\langle B\rangle|$ must be calculated on a per-pixel basis. This can be achieved by relying on the fact that the mean of a vector $x$ is
\begin{equation}
    m_x = \frac{1}{N} \sum_{i=1}^{N}x(i) = \sum_{i=1}^{N}\frac{x(i)}{N}.
\end{equation}
If the mean $m_{x_{N-1}}$ of a subset of $x(i)$ for $i=1,\ldots,N-1$ is already calculated, then the mean of the full ensemble for $i=1,\ldots,N$ is 
\begin{equation}
    m_{x_N} = \frac{x(N) + (N-1) m_{x_{N-1}}}{N}.
    \label{eq:mean}
\end{equation}
Using Equation~\ref{eq:mean}, the effect of a pixel $I(x,y)$ along $C$ on the unipolarity of the foreground (in the $i^{th}$ channel) can be expressed as
\begin{equation}
    \int_{C^+}\frac{1}{N}\sum_{i=1}^{N}\lambda^{\rm{U}+}_i\frac{\frac{|I_i(x,y)|+ a^+_i n^+}{n^++1} - \left|\frac{I_i(x,y) + m^+_i n^+}{n^++1}\right|}{\frac{|I_i(x,y)|+a^+_i n^+}{n^++1}}\mathrm{d}x\mathrm{d}y,
    \label{eq:uniForeground}
\end{equation}
where $a^+_i$ is the absolute mean of $C^+$ (of the $i^{th}$ channel), and $n^+$ is the number of pixels in $C^+$. Note that we are using $\lambda^{\rm{U}+}_i$ to refer to the user-defined weight. Likewise, the unipolarity of the background can be expressed as
\begin{equation}
    \int_{C^-}\frac{1}{N}\sum_{i=1}^{N}\lambda^{\rm{U}-}_i\frac{\frac{|I_i(x,y)|+ a^-_i n^-}{n^-+1} - \left|\frac{I_i(x,y) + m^-_i n^-}{n^-+1}\right|}{\frac{|I_i(x,y)|+a^-_i n^-}{n^-+1}}\mathrm{d}x\mathrm{d}y.
    \label{eq:uniBackground}
\end{equation}

Adding Equations~\ref{eq:uniForeground} and~\ref{eq:uniBackground} to Equation~\ref{eq:acwev} yields our new energy functional:
\begin{equation}
 \begin{split}
  F(\Bar{m}^+,\Bar{m}^-,\Bar{a}^+,\Bar{a}^-,C)=\mu \ell(C)+\int_{C^+}\frac{1}{N}\sum_{i=1}^{N}\lambda^{\rm{H}+}_i|I_i(x,y)-m^+_i|^2\mathrm{d}x\mathrm{d}y\\
  +\int_{C^-}\frac{1}{N}\sum_{i=1}^{N}\lambda^{\rm{H}-}_i|I_i(x,y)-m^-_i|^2\mathrm{d}x\mathrm{d}y\\
  +\int_{C^+}\frac{1}{N}\sum_{i=1}^{N}\lambda^{\rm{U}+}_i\frac{\frac{|I_i(x,y)|+ a^+_i n^+}{n^++1} - \left|\frac{I_i(x,y) + m^+_i n^+}{n^++1}\right|}{\frac{|I_i(x,y)|+a^+_i n^+}{n^++1}}\mathrm{d}x\mathrm{d}y\\
  +\int_{C^-}\frac{1}{N}\sum_{i=1}^{N}\lambda^{\rm{U}-}_i\frac{\frac{|I_i(x,y)|+ a^-_i n^-}{n^-+1} - \left|\frac{I_i(x,y) + m^-_i n^-}{n^-+1}\right|}{\frac{|I_i(x,y)|+a^-_i n^-}{n^-+1}}\mathrm{d}x\mathrm{d}y.
 \end{split}
 \label{eq:acwevu}
\end{equation}
In this formulation, we mirror the structure of the homogeneity-only version of ACWE for a vector-valued image (Equation \ref{eq:acwev}), providing unipolarity constraints for both the foreground and background. This allows the user to prioritize either a unipolar foreground or background in a given channel ($i$) by varying the relationship between $\lambda^{\rm{U}+}_i$ and $\lambda^{\rm{U}-}_i$. Since we wish to enforce unipolarity in the foreground (CH region), and not in the background, the relationship $\lambda^{\rm{U}+}_j/\lambda^{\rm{U}-}_j$ should be $>1$. We refer to the formulation in Equation~\ref{eq:acwevu} as ``QUACK'' (Quantifying Unipolarity via Active Contour Kinetics), and use ``ACWE'' for the methods in \cite{boucheron2016segmentation, grajeda2023}.

\subsection{Effects of Magnetic Unipolarity on Segmentation}
\label{sec:QUACK_full}

Following the pipeline established by \cite{grajeda2023} for the creation of full-scale ($4096\times4096$~pixels) CH segmentations, we first verify that the header information of the Level 1 193~{\AA} EUV image is correct (and update it, if needed, using \verb+aiapy.calibrate.update_pointing+). Next we convert to Level 1.5 images using \verb+aiapy.calibrate.register+ \citep{Barnes2020}. Finally, we correct for limb brightening using the method of \cite{verbeeck2014} and define the initial contour using threshold parameter $\alpha=0.3$. Magnetic field data are prepared by aligning the corresponding HMI magnetogram to the EUV observation using \verb+reproject.reproject_interp+ \citep{repoject_2020}. EUV and HMI data are combined into a two-channel image, with 193~{\AA} EUV data in the first channel ($i=1$) and HMI data in the last ($i=2$). We note that \cite{Ko_2014} calculated unipolarity using an estimate of the radial magnetic field, but we use the LOS field, due to improved stability.

As in \cite{grajeda2023}, off-disk areas in both channels are set to the mean intensity of non-CH regions on a per-channel basis and reset every 10 iterations. QUACK evolution is performed using the same parameters as \cite{grajeda2023} for contour length ($\mu=0$), and for 193~{\AA} homogeneity ($\lambda^{\rm{H}+}_1=1$, $\lambda^{\rm{H}-}_1=1/50$). We set unipolarity parameters $\lambda^{\rm{U}+}_1 = \lambda^{\rm{U}-}_1 = 0$ for the EUV data. For the HMI data, we set homogeneity parameters $\lambda^{\rm{H}+}_2 = \lambda^{\rm{H}-}_2 = 0$, and define the unipolarity parameters as $\lambda^{\rm{U}+}_2=1$, $\lambda^{\rm{U}-}_2=1/50$. This creates a unipolarity ratio $\lambda^{\rm{U}+}_2/\lambda^{\rm{U}-}_2=50$. By setting identical unipolarity and homogeneity parameters we hope to encourage QUACK to provide equal consideration to homogeneity in EUV and magnetic unipolarity. We note, however, that initial experiments performed over CR 2133 at a reduced spatial resolution suggest that if $\lambda^{\rm{H}+}_1=\lambda^{\rm{U}+}_2$ and $\lambda^{\rm{U}+}_2/\lambda^{\rm{U}-}_2\geq10$, the specific choice of $\lambda^{\rm{U}-}_2$ has negligible effect.

\begin{figure}
    \centering
    \subfloat[Intersection Over Union]{\includegraphics[trim=0in 1.2in 0in .7in,clip,width=.49\textwidth]{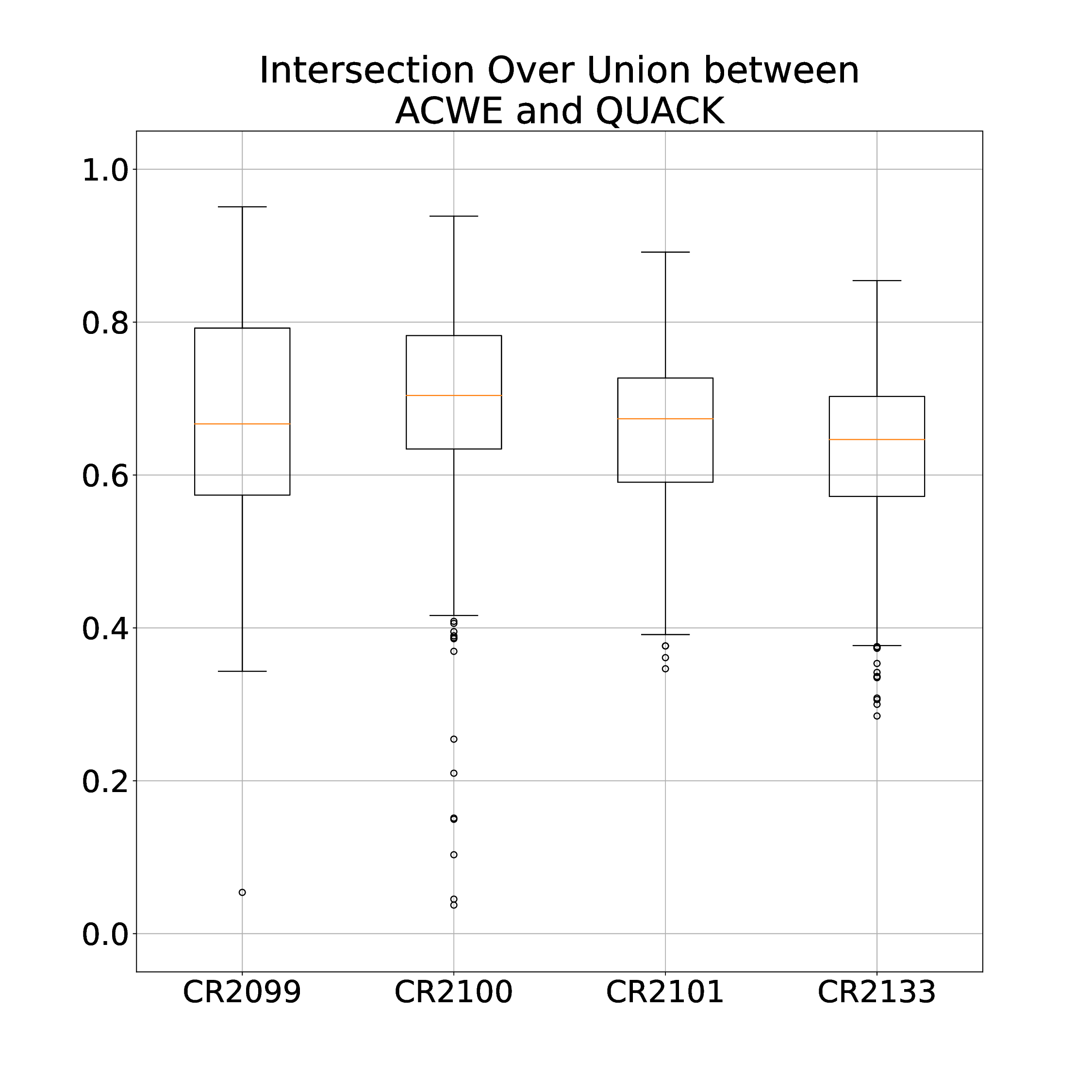}}~~
    \subfloat[Structural Similarity]{\includegraphics[trim=0in 1.2in 0in .7in,clip,width=.49\textwidth]{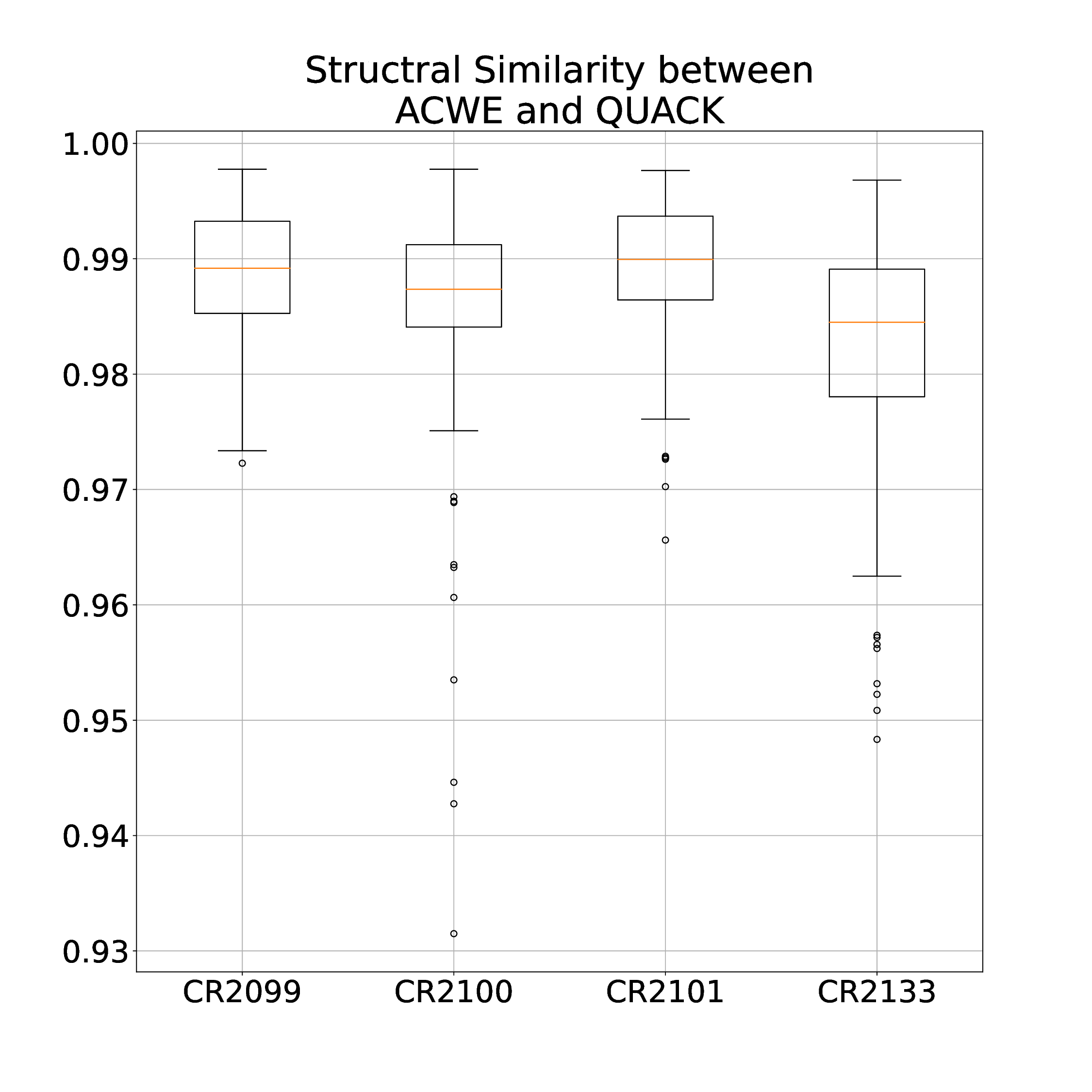}}\\
    \subfloat[Global Consistency Error]{\includegraphics[trim=0in 1.2in 0in .7in,clip,width=.49\textwidth]{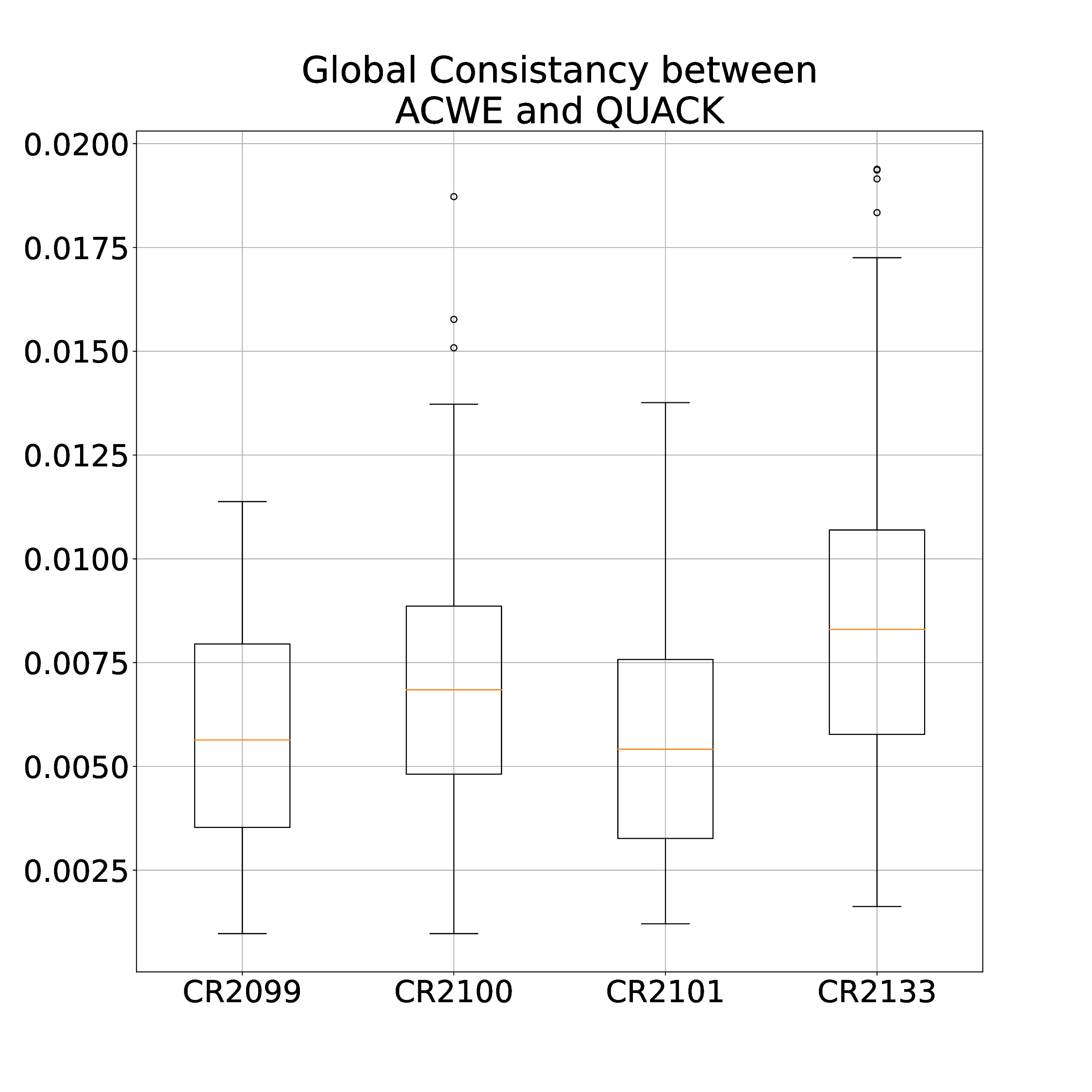}}~~
    \subfloat[Local Consistency Error]{\includegraphics[trim=0in 1.2in 0in .7in,clip,width=.49\textwidth]{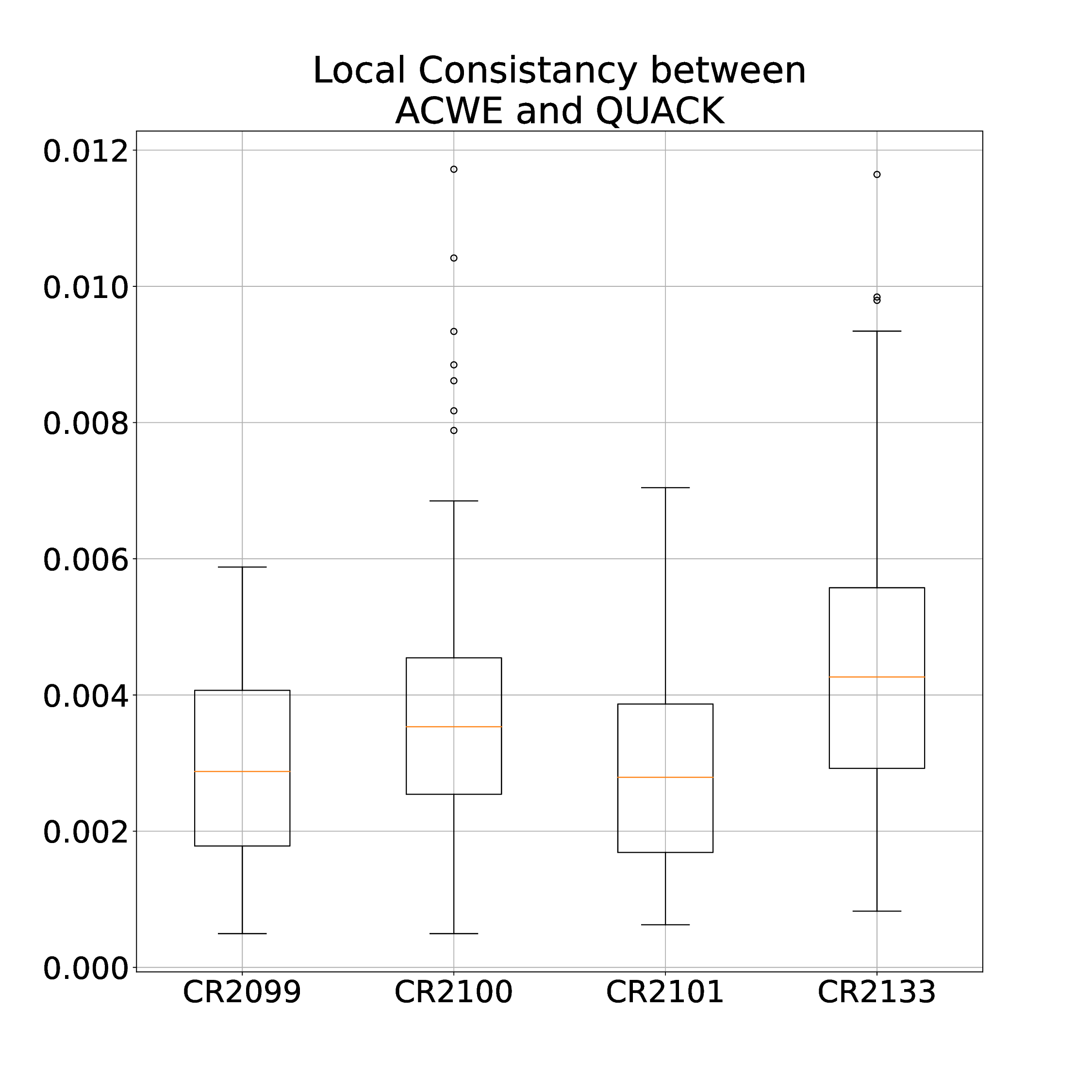}}\\
    \caption{Comparison between full-scale segmentation results using only 193~{\AA} data (ACWE), and using both 193~{\AA} and HMI magnetogram data (QUACK), organized by CR. The box outlines the range between Q1 and Q3, with the median value in orange. The whiskers show 1.5 times the interquartile range. Outliers are marked with circles.}
    \label{fig:MagBoxPlots}
\end{figure}

\begin{figure}
    \centering
    \subfloat[IOU:0.28476]{\includegraphics[trim=0in .45in 0in 0in,clip,width=\textwidth]{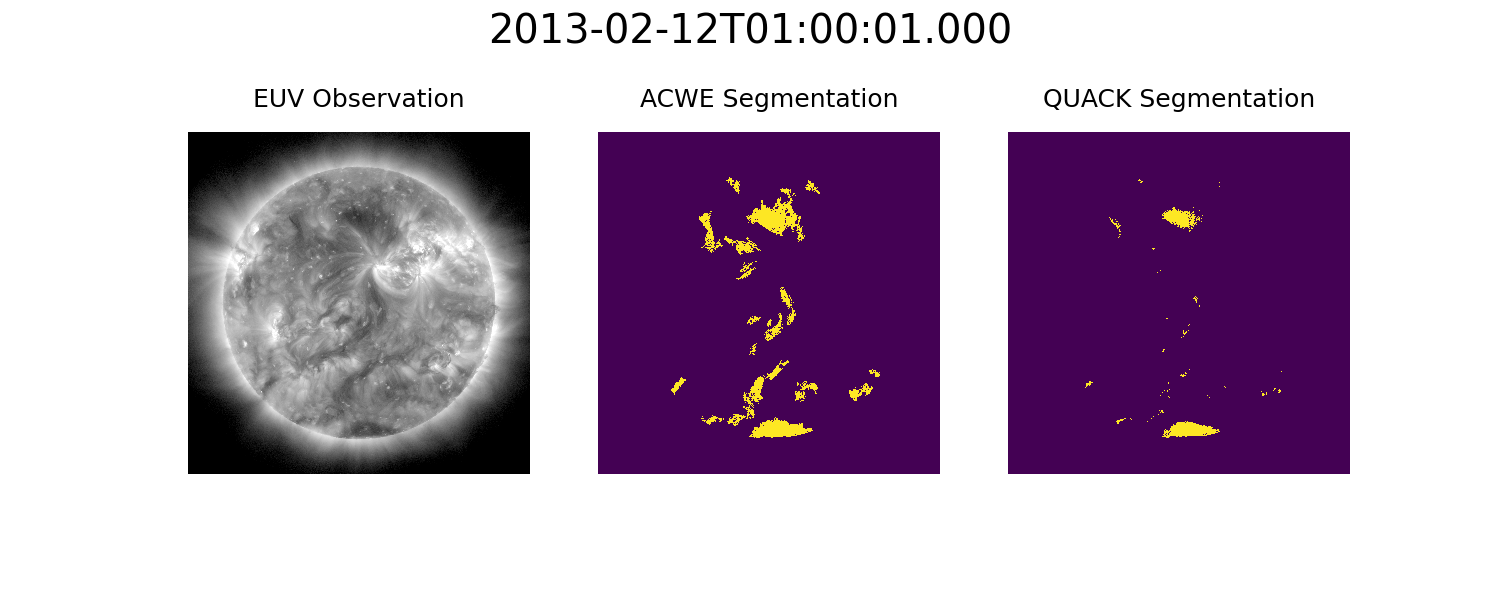}}\\
    \subfloat[IOU:0.30002]{\includegraphics[trim=0in .45in 0in 0in,clip,width=\textwidth]{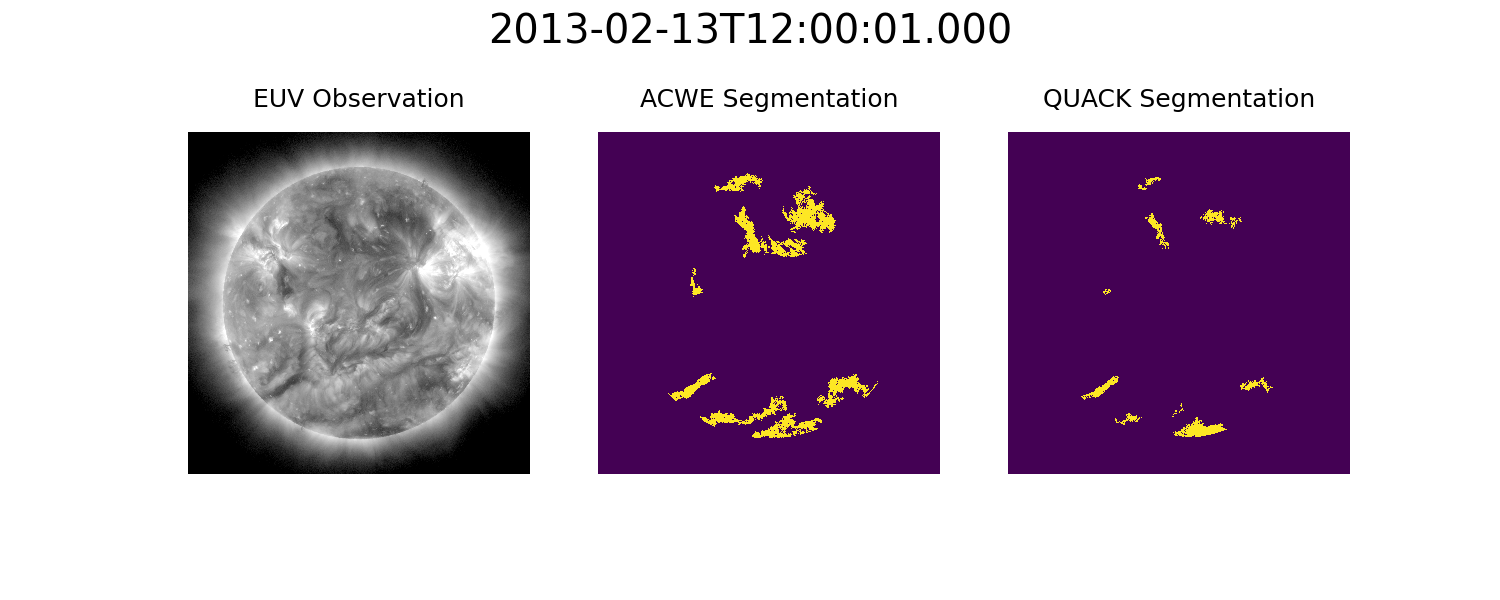}}\\
    \subfloat[IOU:0.30628]{\includegraphics[trim=0in .45in 0in 0in,clip,width=\textwidth]{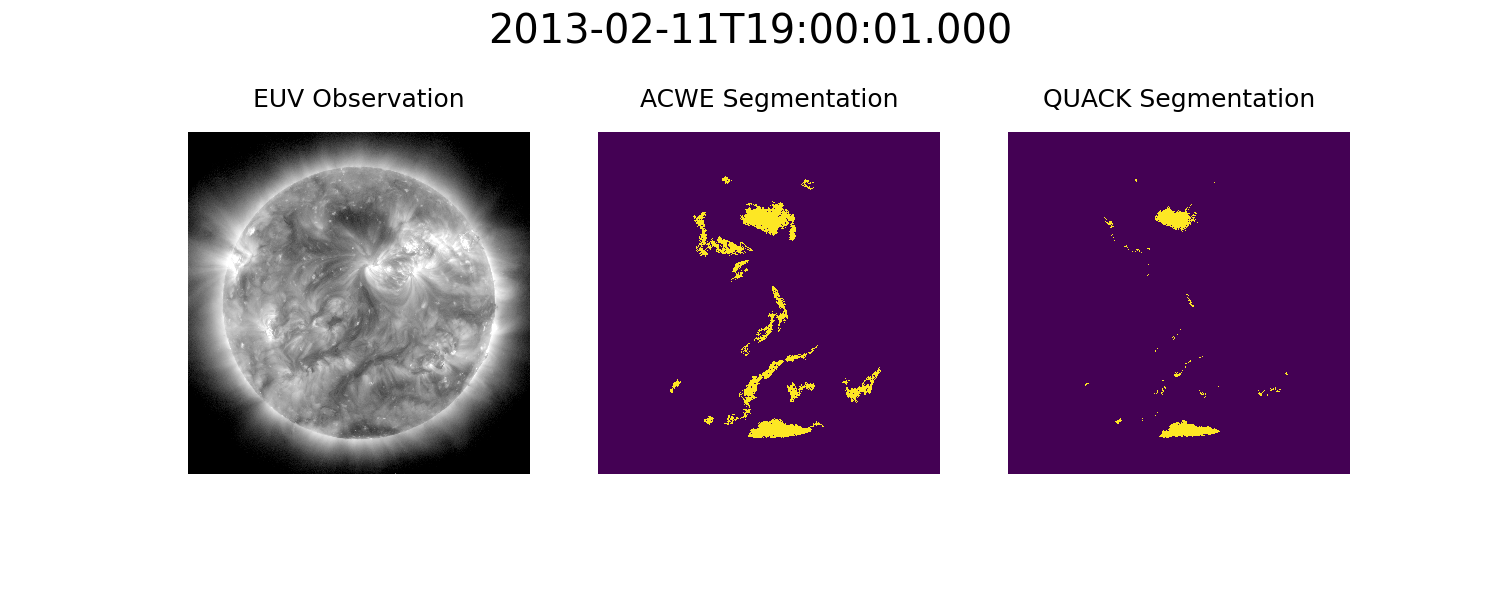}}\\
    \caption{Examples of full-scale segmentations using only 193~{\AA} (ACWE, center), and using both 193~{\AA} and HMI (QUACK, right) over CR 2133. The leftmost image is the 193~{\AA} observation. The title of each figure is the record time for the EUV data.}
    \label{fig:MagSamplesCR2133}
\end{figure}

The IOU, SSIM, GCE, and LCE between the full-scale 193~{\AA} ACWE segmentations of \cite{grajeda2023} and QUACK, organized by CR, are presented in Figure~\ref{fig:MagBoxPlots}. We note that, across all four metrics, CR 2133 displays the largest median discrepancy. Given that filaments are most prevalent in CR 2133, this is consistent with the expectation that the magnetic unipolarity parameters are reducing filament contamination. Visual inspection of the segmentation pairs with the lowest IOU for CR2133, of which three samples are provided in Figure~\ref{fig:MagSamplesCR2133}, confirms this speculation by showing that filament regions, while still present, are consistently diminished to a small region surrounding the initial seed.

\begin{figure}
    \centering
    \subfloat[IOU:0.05399]{\includegraphics[trim=0in .45in 0in 0in,clip,width=\textwidth]{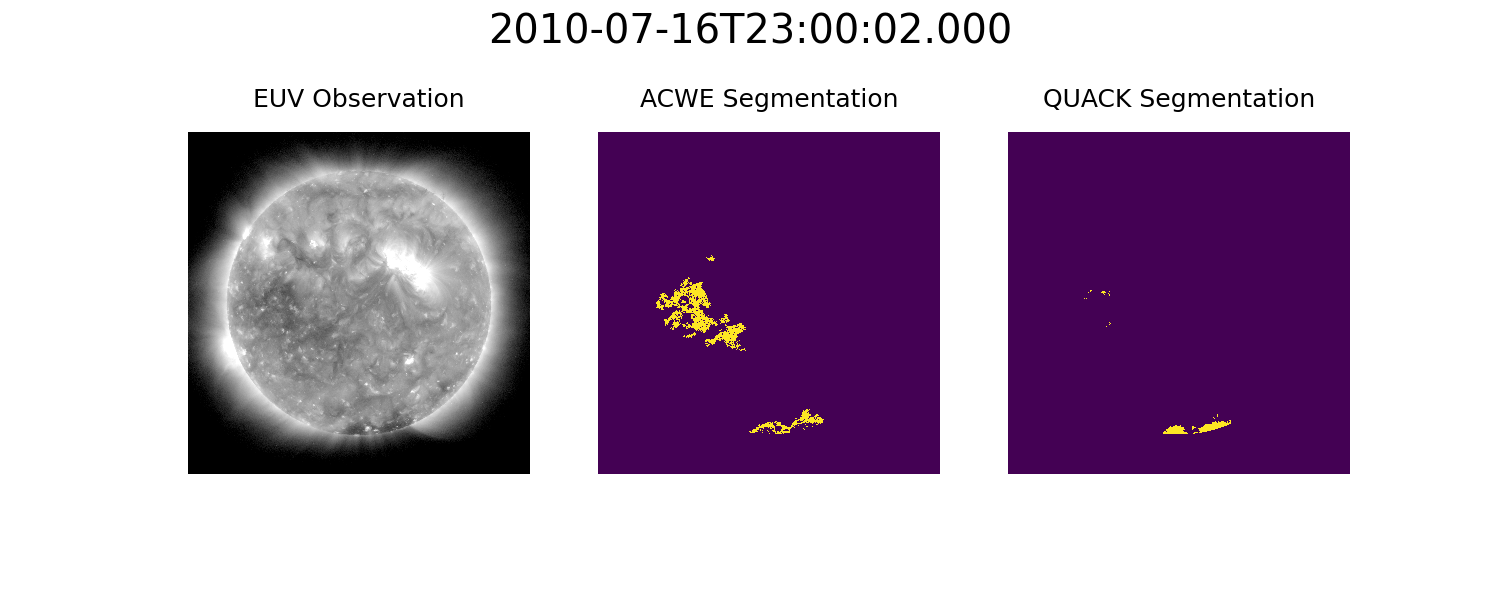}}\\
    \subfloat[IOU:0.34324]{\includegraphics[trim=0in .45in 0in 0in,clip,width=\textwidth]{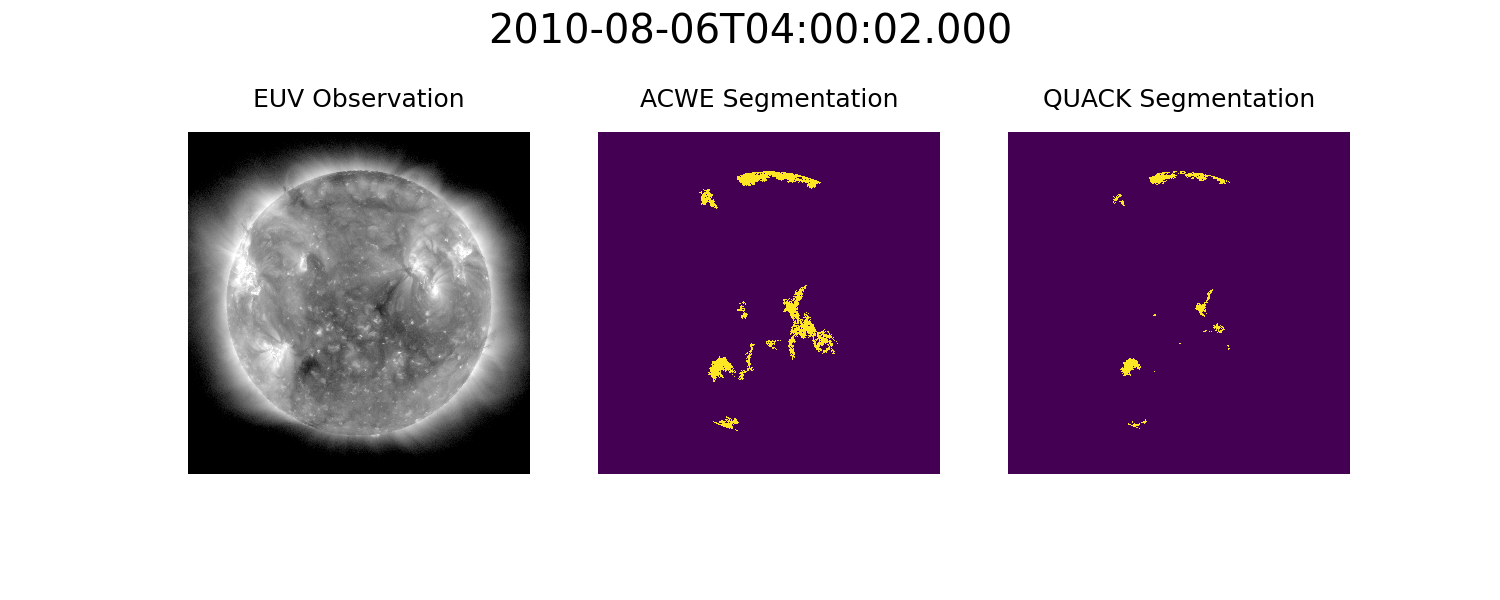}}\\
    \subfloat[IOU:0.34519]{\includegraphics[trim=0in .45in 0in 0in,clip,width=\textwidth]{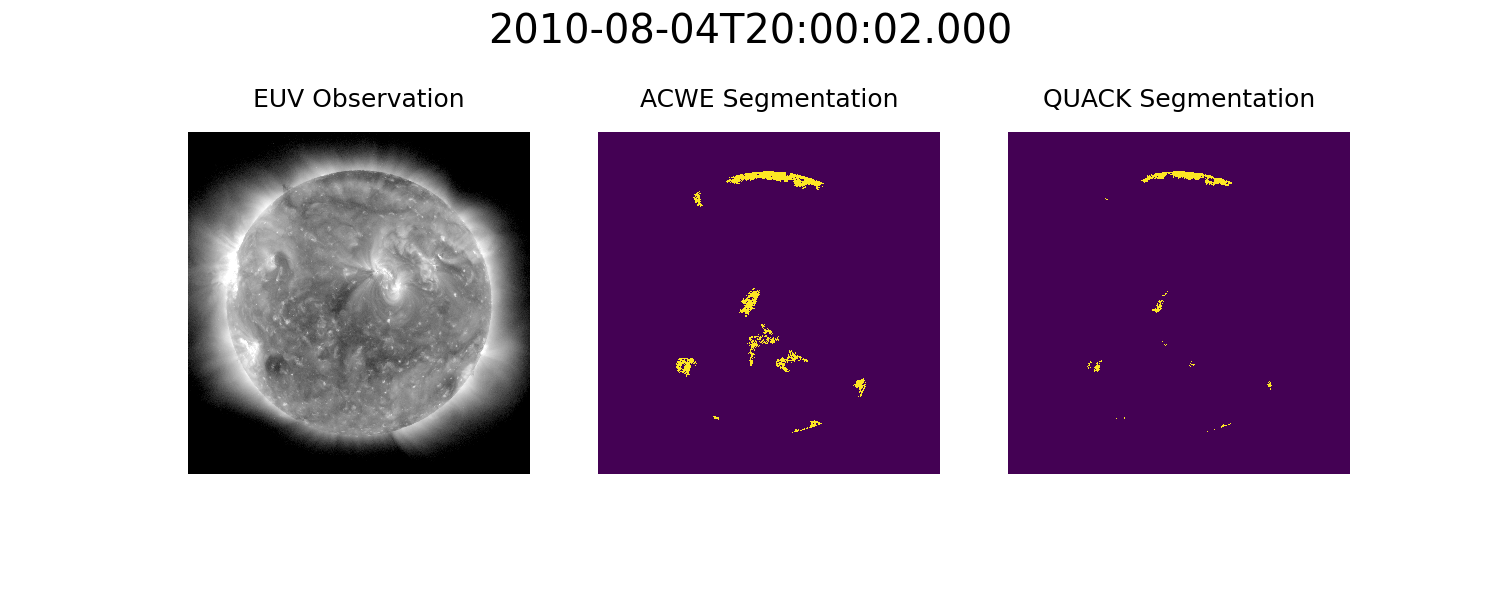}}\\
    \caption{Examples of full-scale segmentations using only 193~{\AA} (ACWE, center), and using both 193~{\AA} and HMI (QUACK, right) over CR 2099. The leftmost image is the 193~{\AA} observation. The title of each figure is the record time for the EUV data.}
    \label{fig:MagSamplesCR2099}
\end{figure}

Referring to Figure~\ref{fig:MagBoxPlots}, we note that that CR 2099 shows a large range between Q1 and Q3 in IOU, GCE, and LCE, but not in SSIM. We also note that CR 2099 shows the second largest median discrepancy in IOU, but not in the remaining three metrics. These results are consistent with a change in overall CH area while retaining the overall form or shape of the identified CHs. \cite{grajeda2023} note that CR 2099 had the largest number of segmentations that changed from targeting CHs to QS.  This may suggest that QUACK is reducing the growth of $C^+$ into QS regions, which is another motivation for inclusion of magnetic field into evolution rather than just post-hoc processing. Visual inspection of the segmentation pairs with the lowest IOU for CR 2099, of which three samples are provided in Figure~\ref{fig:MagSamplesCR2099}, also confirms this.

In summary, incorporating magnetic field information, via a unipolarity metric operating on HMI data, appears to have negligible impact on segmentation of CHs.  At the same time, it appears to minimize area of detected filaments and alleviate many cases of change of target.  We return to these results in Section~\ref{sec:Seed} after first considering effects of spatial resolution on segmentation accuracy.

\subsection{Effects of Spatial Resolution on Segmentation Accuracy}
\label{sec:QUACK_eight}

The iterative nature of ACWE results in segmentations at full-scale taking several minutes, while reducing the image to one-eighth-scale resolution ($512\times512$~pixels) reduces segmentation time to seconds \citep{grajeda2023}. To determine the viability of using one-eighth-scale resolution images for QUACK segmentations, we replicate the pipeline of \cite{grajeda2023} for reduced scale segmentations. We reduce the 193~{\AA} image and corresponding magnetogram to $512\times512$~pixels using scikit-image function \verb|skimage.transform.resize| \citep{scikit-image} after aligning the observations, but before correcting for limb brightening and before defining the seed. We evolve the contour on the decimated images, using the same weights as Section~\ref{sec:QUACK_full}. After each segmentation has converged, it is scaled back to $4096\times4096$~pixels and compared with the QUACK segmentations of Section~\ref{sec:QUACK_full}.

\begin{figure}
    \centering
    \subfloat[Intersection Over Union]{\includegraphics[trim=0in 1.2in 0in 1.3in,clip,width=0.45\textwidth]{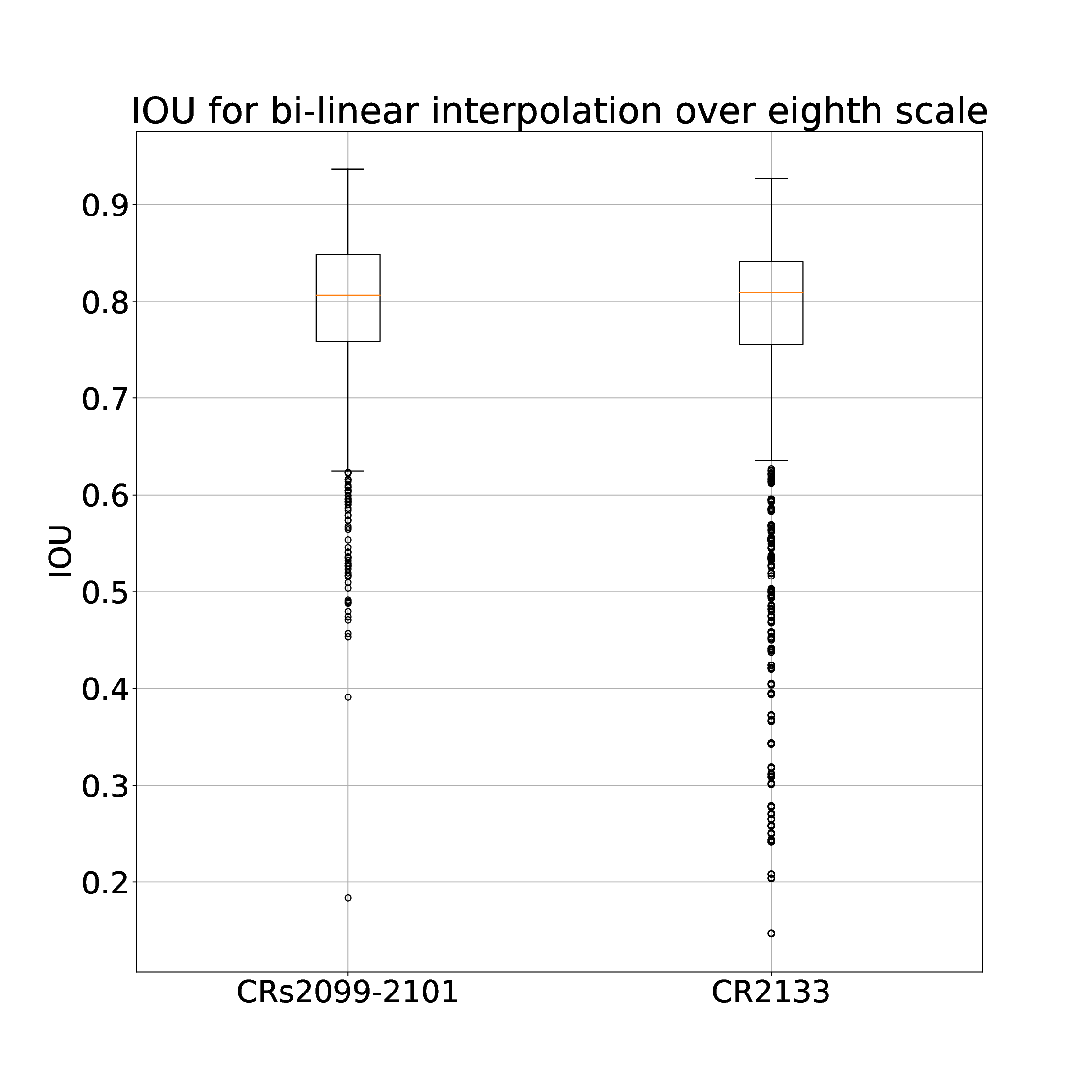}}~
    \subfloat[Structural Similarity]{\includegraphics[trim=0in 1.2in 0in 1.3in,clip,width=0.45\textwidth]{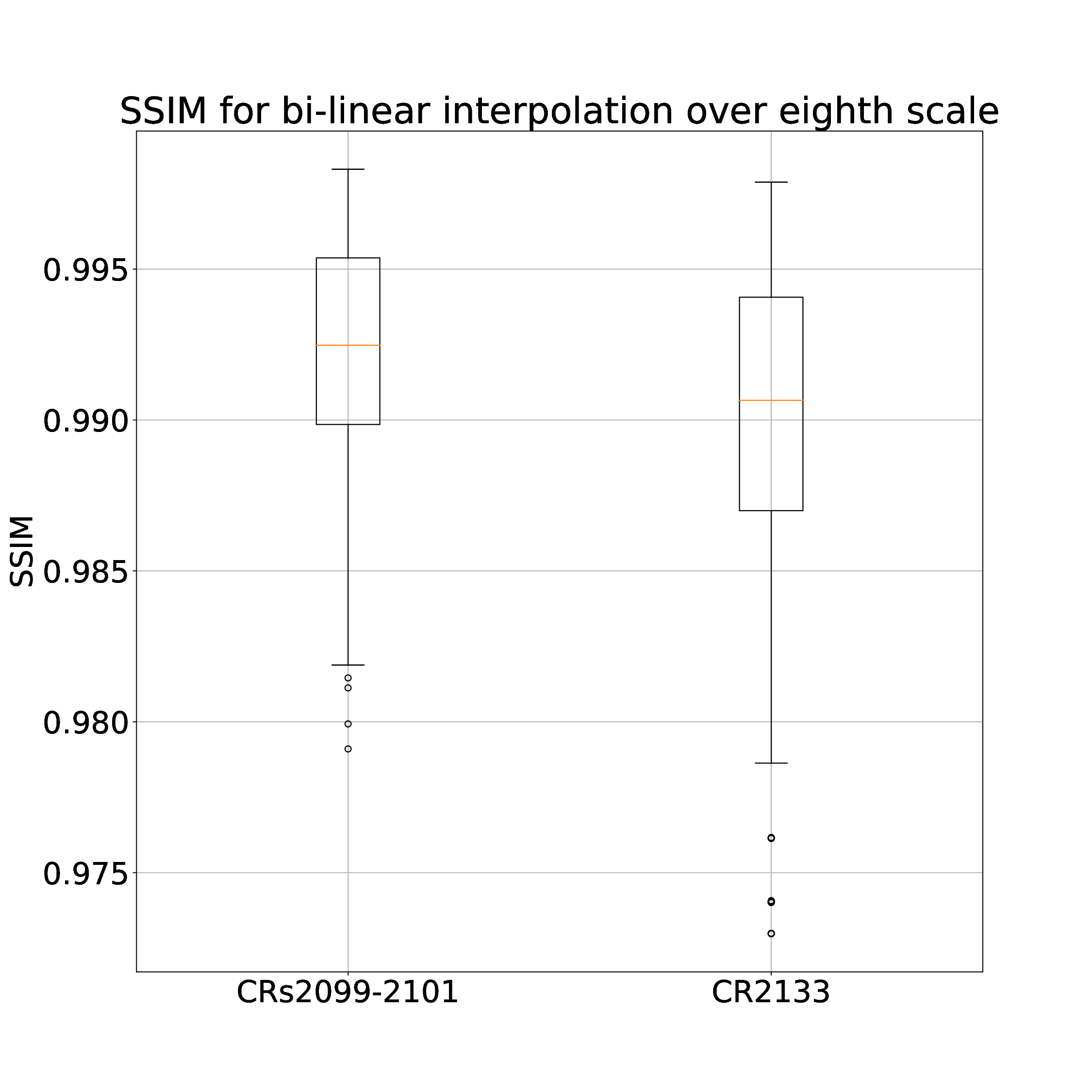}}\\
    \subfloat[Global Consistency Error]{\includegraphics[trim=0in 1.2in 0in 1.3in,clip,width=0.45\textwidth]{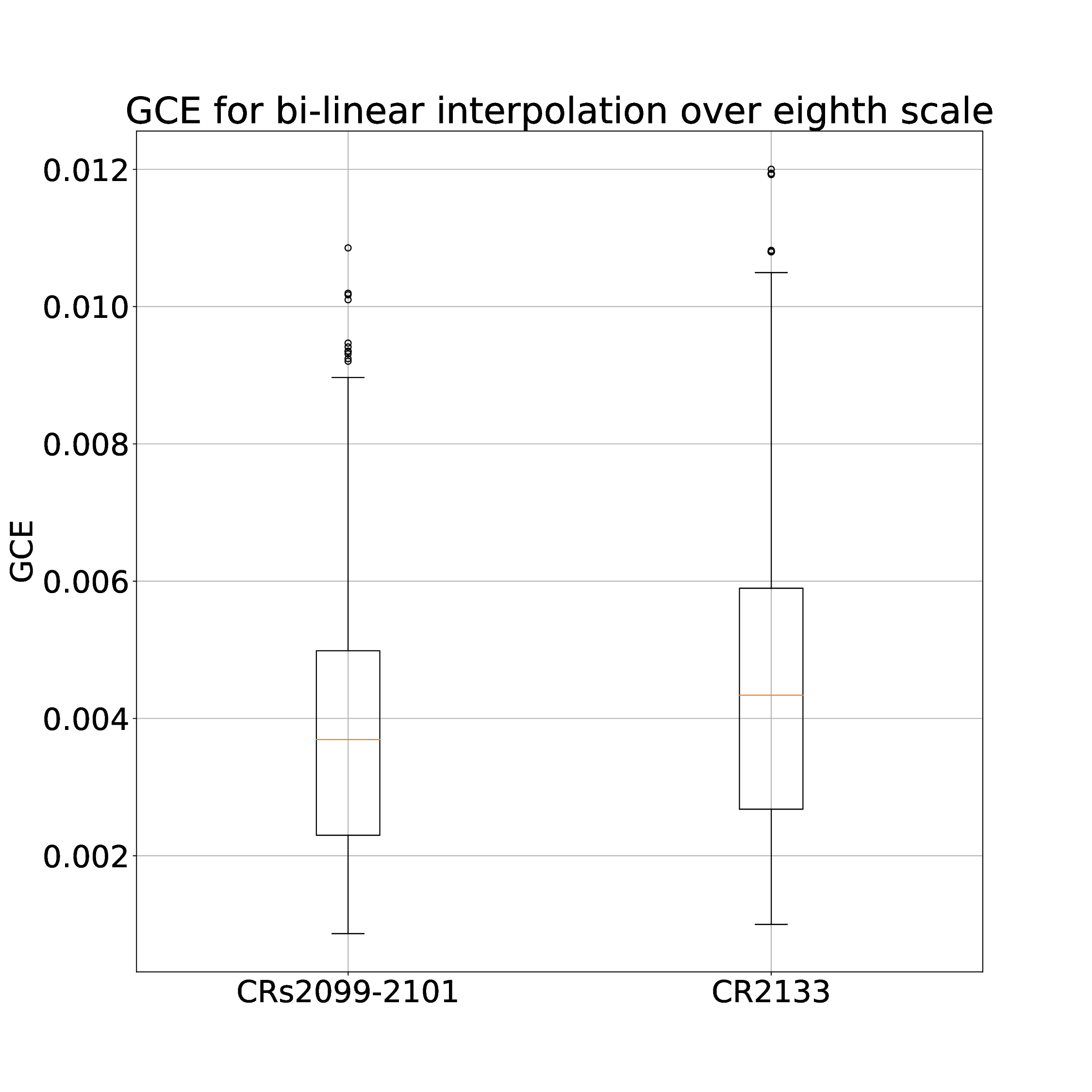}}~
    \subfloat[Local Consistency Error]{\includegraphics[trim=0in 1.2in 0in 1.3in,clip,width=0.45\textwidth]{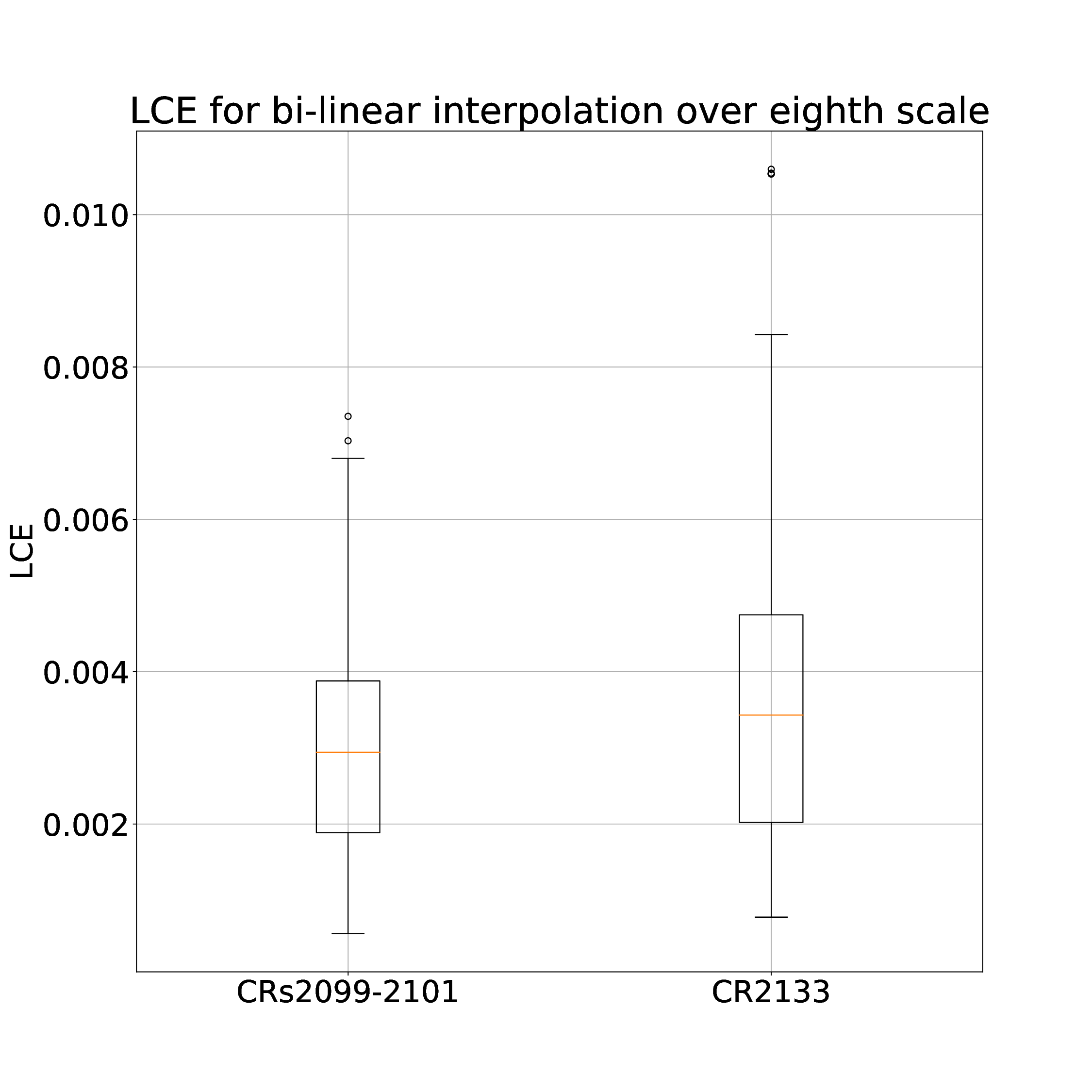}}\\
    \caption{Comparison between full-scale and one-eighth-scale QUACK segmentations for CRs 2099-2010 (left) and CR 2133 (right). The box outlines the range between Q1 and Q3, with the median value in orange. The whiskers show 1.5 times the interquartile range. Outliers are marked with circles.}
    \label{fig:Scale}
\end{figure}

The effects of reducing spatial resolution on segmentation similarity are summarized in Figure~\ref{fig:Scale}. In this figure, we provide similarity for bi-linear interpolation, noting that five additional interpolation methods (nearest-neighbor, bi-quadratic, bi-cubic, bi-quartic, and bi-quintic) yielded similar results. Results for CRs 2099-2101 (left in each plot) are consistent with \cite{grajeda2023}, and results for CR 2133 (right in each plot) are improved. This suggests that decimation to one-eighth-scale is a viable means to reduce algorithm runtime despite (and in some cases, because of) the decimated magnetogram data.

\begin{figure}
    \centering
    \includegraphics[trim=0in 2.5in 0in 0in clip, width=\textwidth]{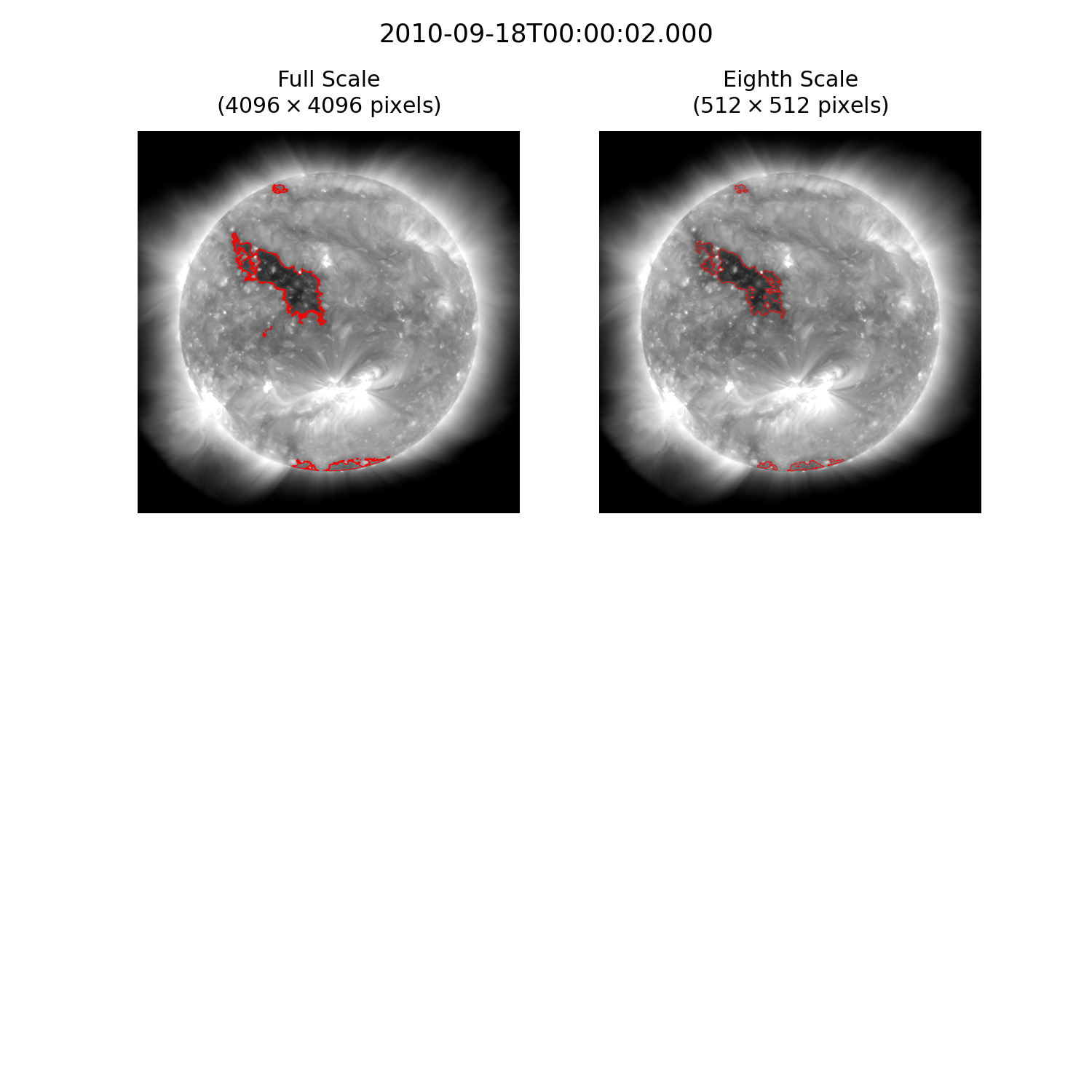}
    \caption{Example of full-scale and one-eighth-scale QUACK segmentations. The title of the figure is the record time of the EUV data.}
    \label{fig:ScaleSample}
\end{figure}

Visual inspection revealed the same three discrepancies between full- and reduced-resolution segmentations as in \cite{grajeda2023}: 1) absence of smaller regions due to downsampling removing dark pixels that formed the initial seed, 2) presence or absence of spurious bright regions at different scales, and 3) reduced fidelity along the contour boundary due to the lower spatial resolution. All three effects are demonstrated in Figure~\ref{fig:ScaleSample}. No new discrepancies appear to have been added by introducing spatially downsampled magnetogram data. 

Absence of smaller regions due to downsampling is especially prevalent in the case of filaments, which, due to their thin structure, are more likely to lose dark pixels that form the initial seed when downsampled. This explains the larger discrepancy \cite{grajeda2023} found when rescaling CR 2133. In particular, the fact that adding magnetogram data constrains filament evolution at full-scale ensures higher similarity between filament-absent one-eighth-scale segmentations and filament-present full-scale segmentations.

\begin{figure}
    \centering
    \subfloat[Intersection Over Union]{\includegraphics[trim=0in 1.2in 0in .7in,clip,width=0.45\textwidth]{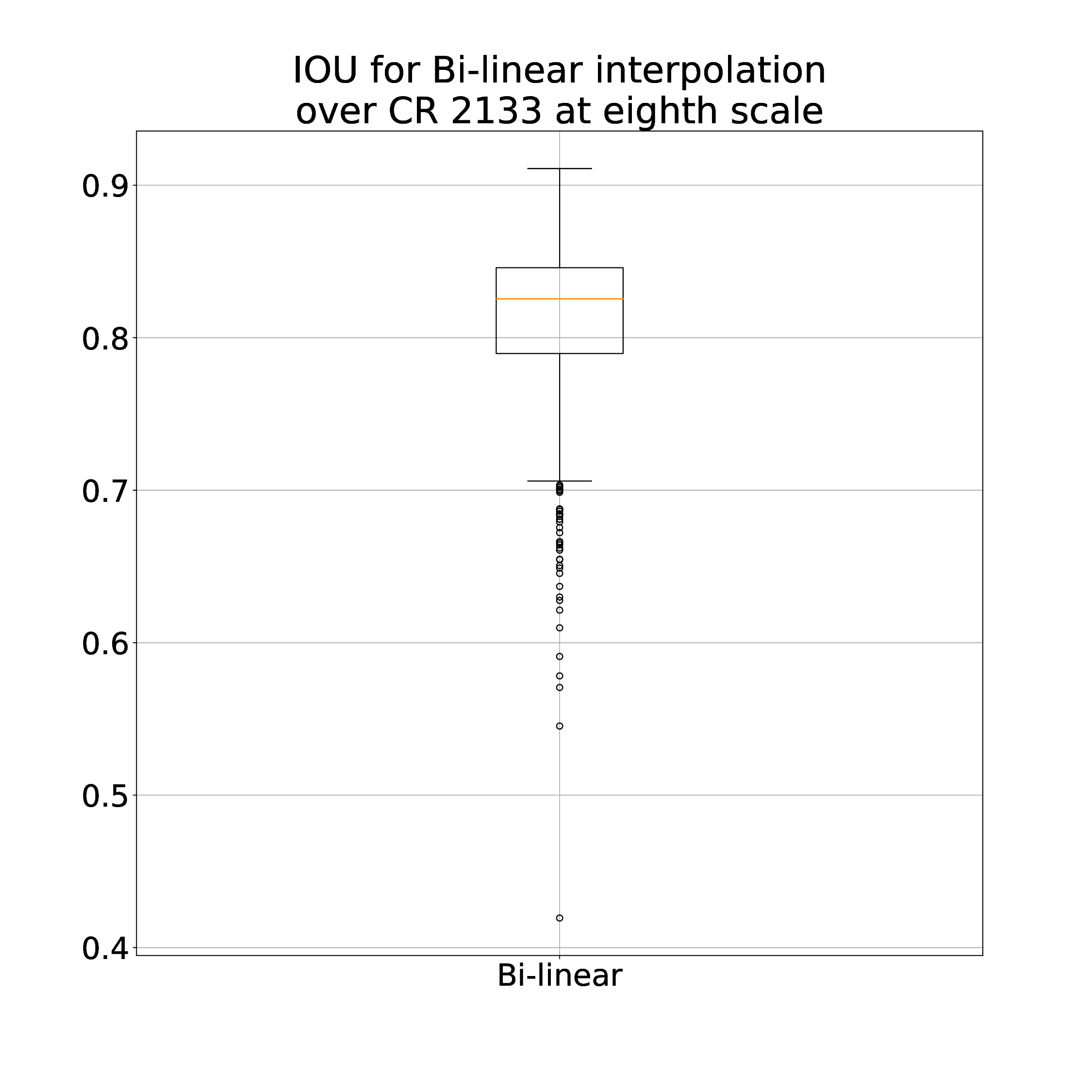}}~
    \subfloat[Structural Similarity]{\includegraphics[trim=0in 1.2in 0in .7in,clip,width=0.45\textwidth]{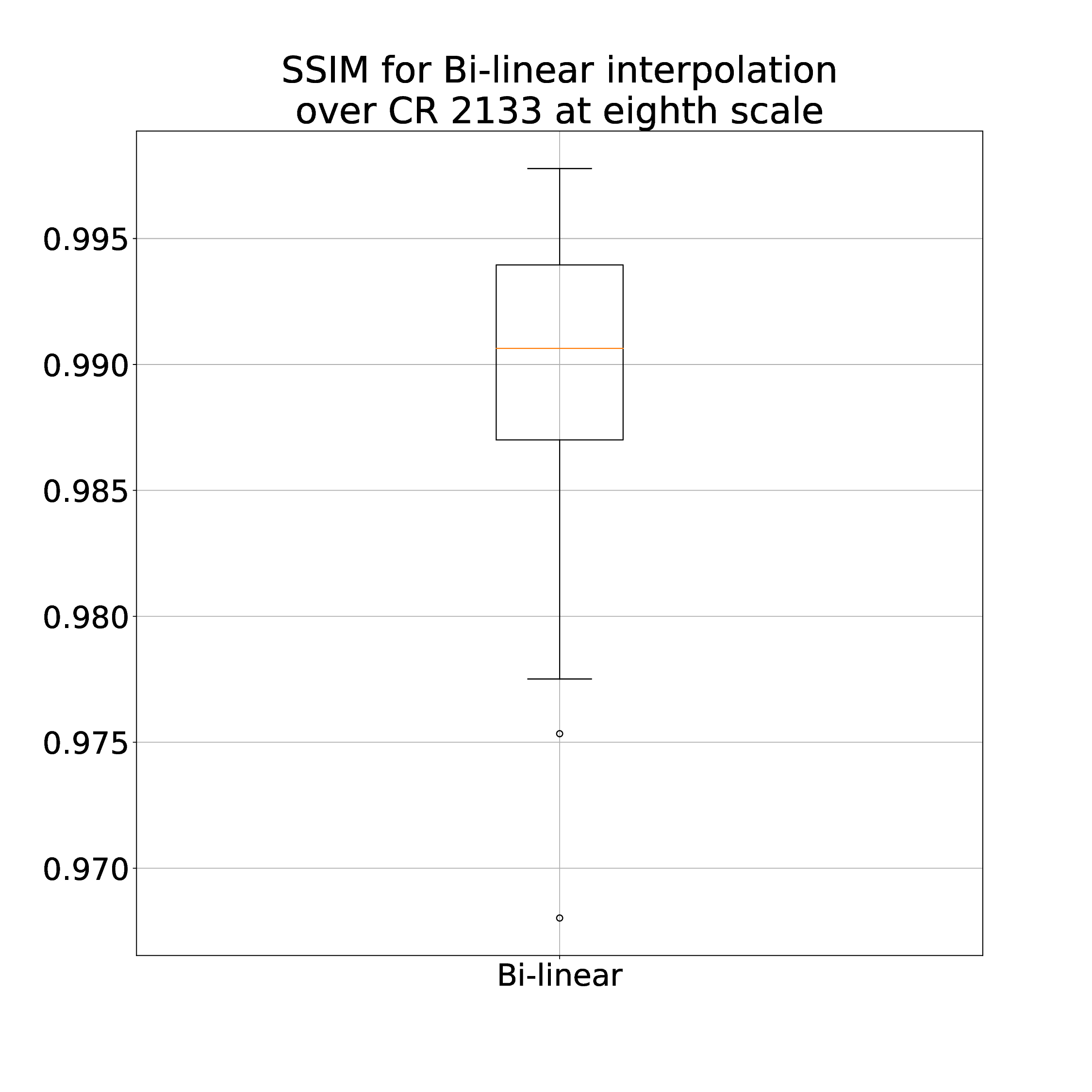}}\\
    \subfloat[Global Consistency Error]{\includegraphics[trim=0in 1.2in 0in .7in,clip,width=.45\textwidth]{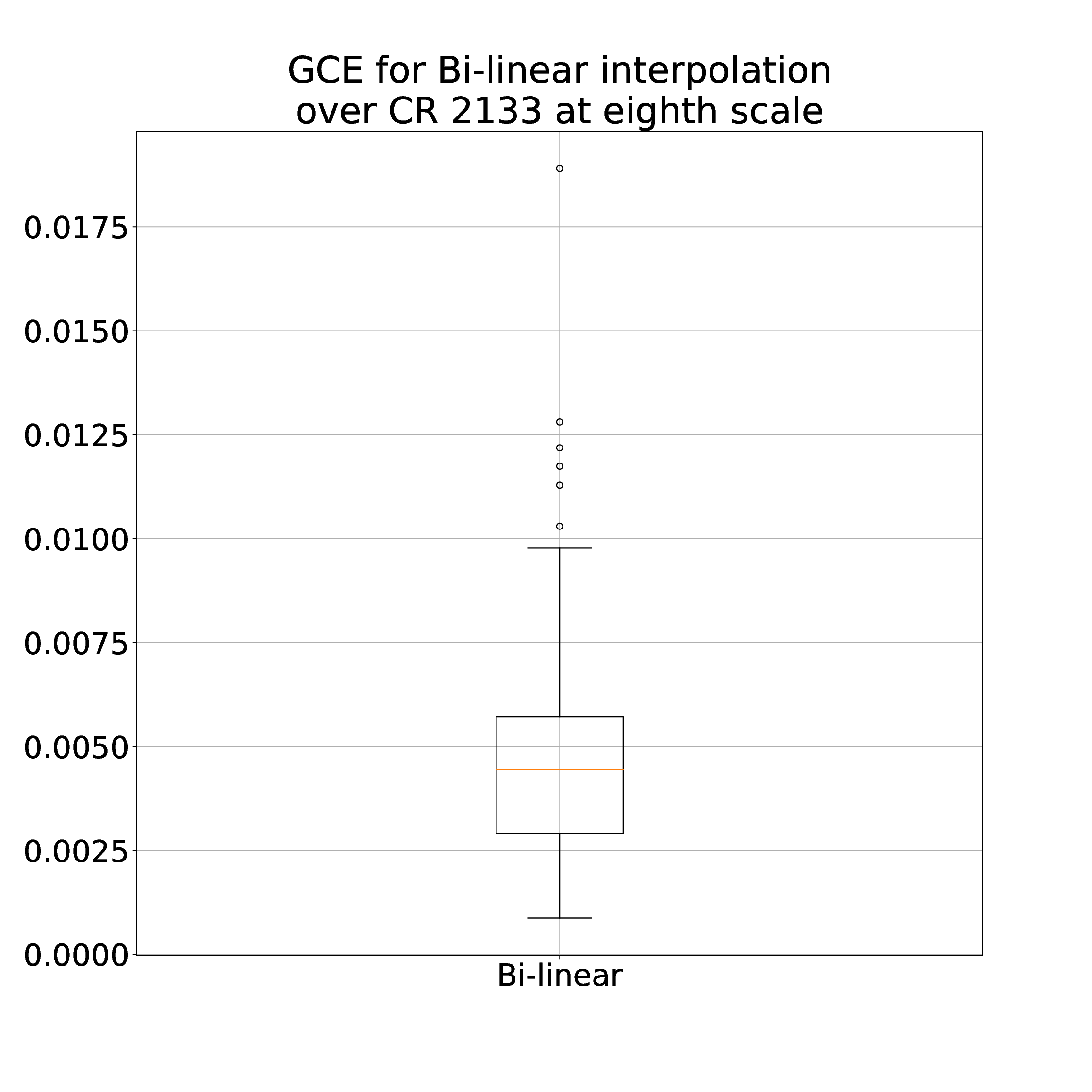}}~
    \subfloat[Local Consistency Error]{\includegraphics[trim=0in 1.2in 0in .7in,clip,width=.45\textwidth]{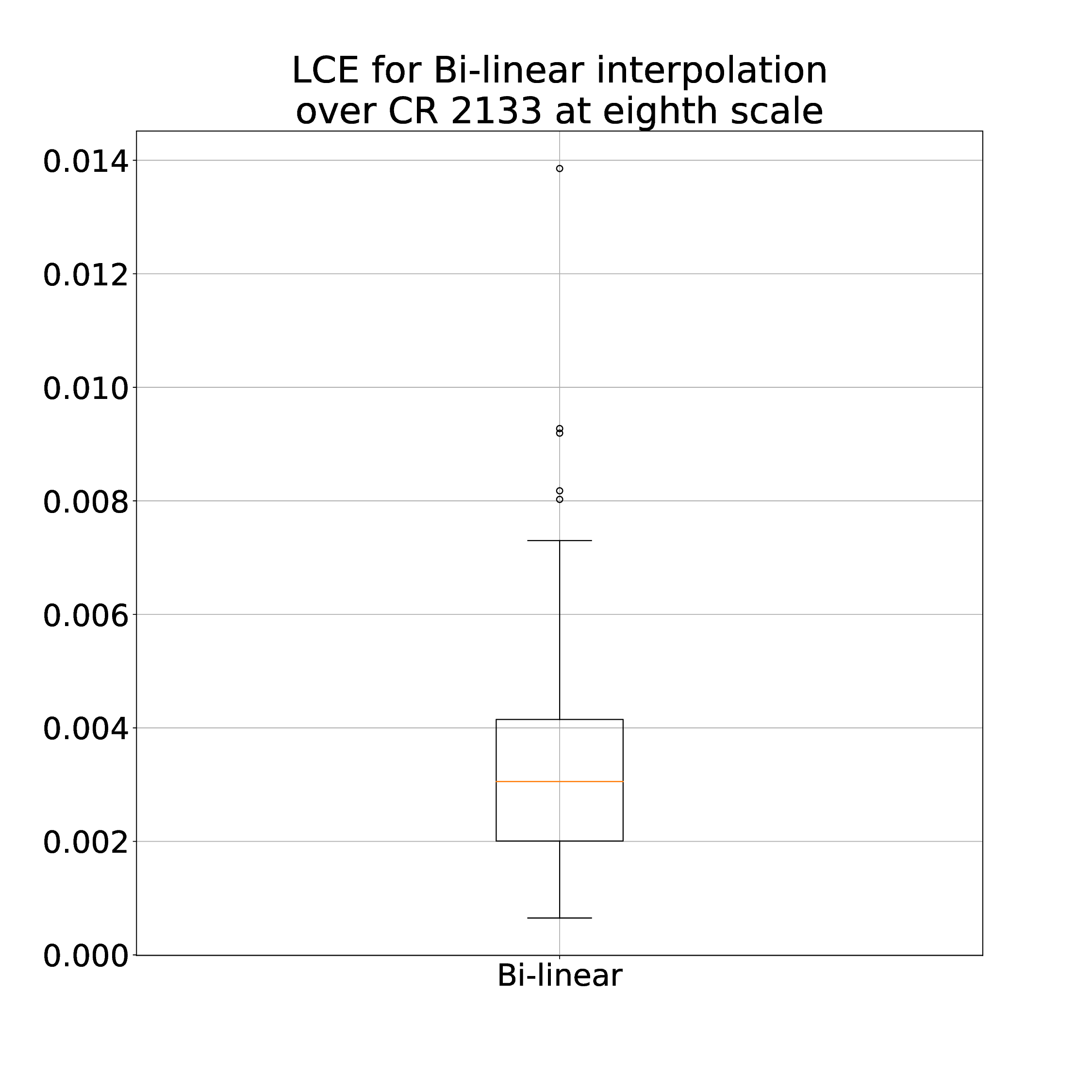}}\\
    \caption{Comparison between full-scale and one-eighth-scale QUACK segmentations for CR 2133 when the full-scale seed is utilized to generate the one-eighth-scale segmentation. The box outlines the range between Q1 and Q3, with the median value in orange. The whiskers show 1.5 times the interquartile range. Outliers are marked with circles.}
    \label{fig:Scale_CR2133T}
\end{figure}

\begin{figure}
    \centering
    \subfloat
    {\includegraphics[trim=0in .45in 0in 0in,clip,width=\textwidth]{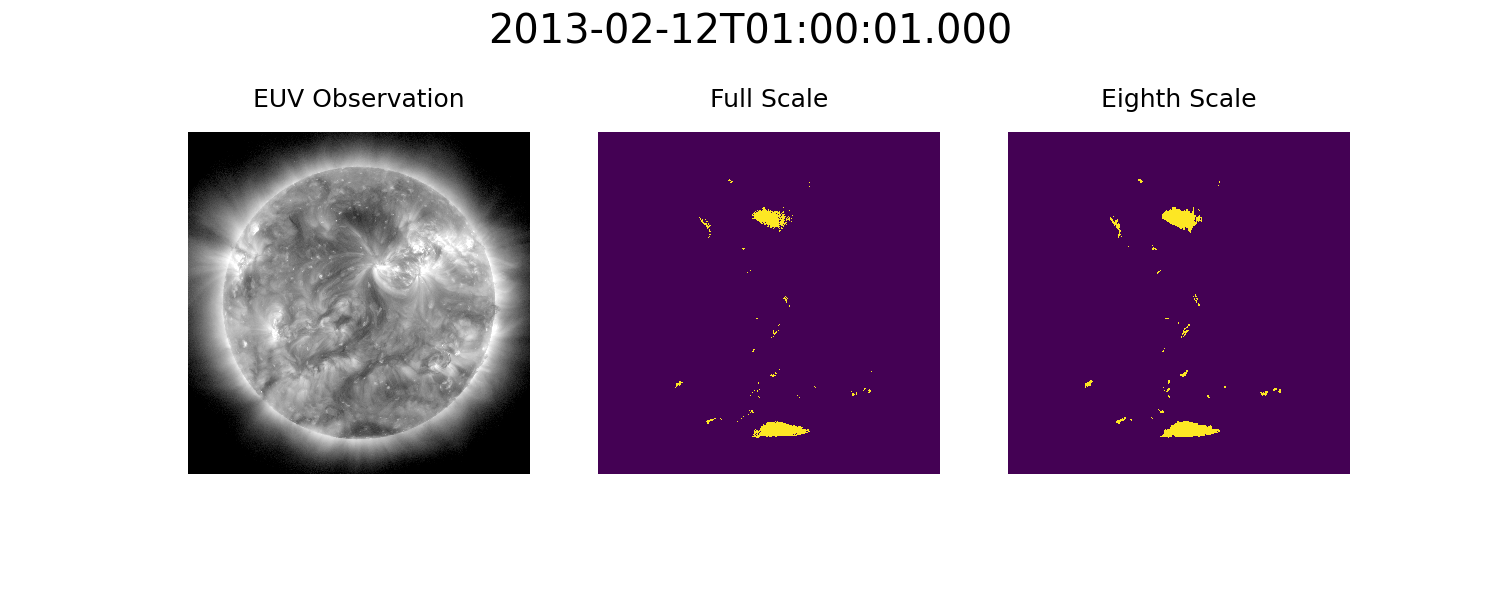}}\\
    \subfloat
    {\includegraphics[trim=0in .45in 0in 0in,clip,width=\textwidth]{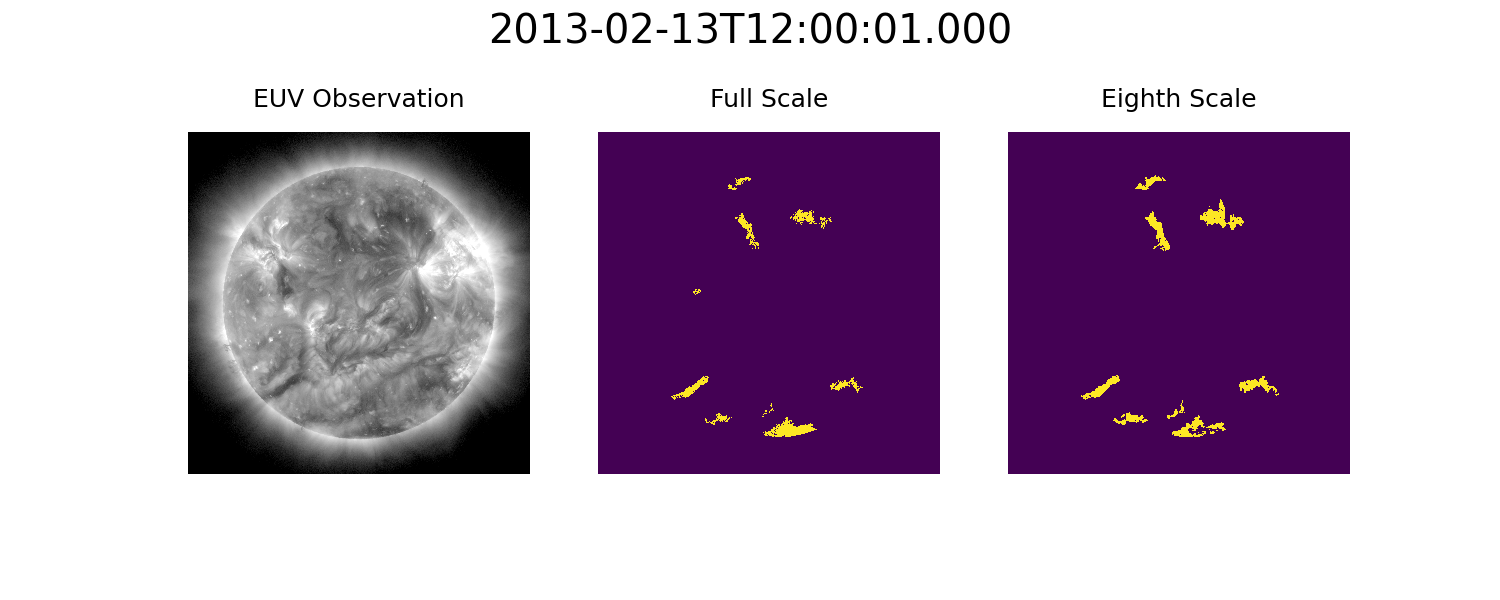}}\\
    \subfloat
    {\includegraphics[trim=0in .45in 0in 0in,clip,width=\textwidth]{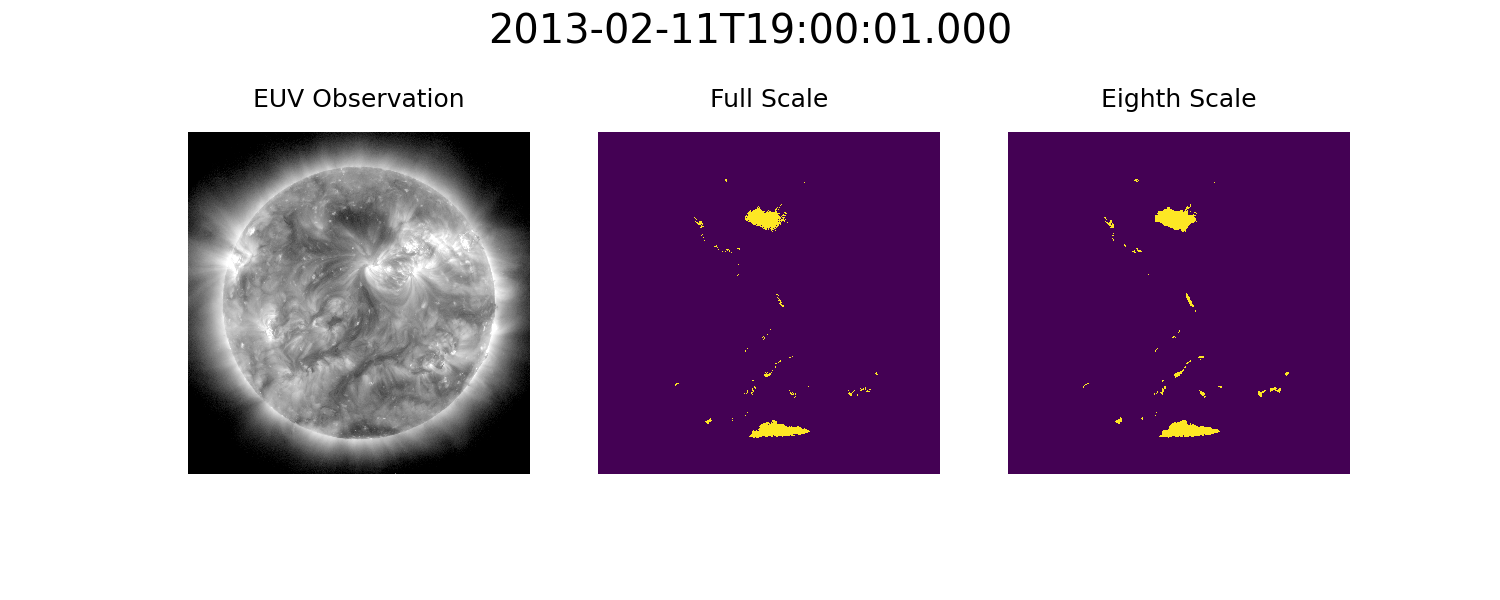}}\\
    \caption{Effects of seed transfer on QUACK segmentations generated at one-eighth-scale (right) compared to the corresponding full-scale segmentations (center). In each figure the leftmost image is the 193~{\AA} observation and the title is the record time for the EUV data.}
    \label{fig:MagScaleSamplesCR2133}
\end{figure}

Due to the aforementioned absence of some filaments in CR 2133 at one-eighth-scale, we additionally studied the effects of transferring the seed (initial contour) generated at full-scale resolution to the one-eighth scale resolution. This is done to ensure that all filaments present in full-scale segmentations are also in the one-eighth scale segmentations, and therefore to verify that spatially-decimated magnetogram data still effectively constrain CH segmentations. Due to differing spatial resolutions ($4096\times4096$ versus $512\times512$~pixels), the full-scale seed must be downsampled to the new resolution. Since every pixel at the one-eighth-scale resolution represents an $8\times8$~pixel patch of the original image, we apply dilation with a $4\times4$~pixel square using scikit-image function \verb|skimage.morphology.dilation| \citep{scikit-image} prior to downsampling the initial seed. This ensures that a region represented with even only one pixel in the full-scale seed is retained at the one-eighth-scale resolution. The effects of the seed transfer are summarized in Figure~\ref{fig:Scale_CR2133T}. We note that seed transfer had minimal effect on SSIM, GCE, and LCE, but further improved IOU, suggesting that the added filaments were constrained in the same manner as before. This is confirmed through a visual inspection of the data. For reference, Figure~\ref{fig:MagScaleSamplesCR2133} provides full-scale and one-eighth-scale segmentations generated utilizing the seed transfer process for the same examples in Figure~\ref{fig:MagSamplesCR2133}. Transferring the seed for the remaining three CRs had no appreciable effect.

In summary, reducing the resolution of HMI magnetograms via standard image resizing procedures does not appear to affect segmentation performance.  We note that this image decimation does not take into account magnetic flux conservation (i.e., correction for differing areas of pixels across the disk), but appears to retain statistics relevant to measures of unipolarity.  Additionally, we find that a loss of seeds (individual regions in the initial contour) related to filaments does not appear to be the largest contributing source of this performance.  Rather, it appears that reduced-resolution HMI data retains relevant information for constraining evolution of the contour, reducing the contribution of filaments.  Since filaments are still present, albeit with reduced area, we now consider incorporation of magnetic field information into the seeding process to further reduce contamination.  This reinforces the importance of incorporating magnetic field information throughout CH segmentation rather than as a post-hoc process.

\section{Incorporation of Magnetic Field Data into the Seeding Process}
\label{sec:Seed}

Due to reduced filament contamination, improved runtime, and minimal effect on segmentation, we operate at one-eighth scale for all steps in QUACK. We recognize, however, that while this improves the quality of the segmentations, filament contamination, though minimized, persists. This section, therefore, explores the use of magnetic field data in initial seeding to further reduce contamination.

\subsection{Complications with Estimating Unipolarity from Initial Seeds}
\label{sec:SeedSplitAndOverfit}

Predicting the unipolarity of a final region through direct evaluation of the seed, regardless of the scale at which seeding occurred, is not possible due to two issues. First, multiple seeds (individual, spatially unconnected regions in the seed) often contribute to a single CH (these seeds merge into the final region during contour evolution). Second, these seeds are often small regions vulnerable to the effects of small-scale statistics, which results in erroneous estimations of unipolarity at the seeding stage. In order to demonstrate these two issues, we briefly return to ACWE (EUV-only) segmentations for the following two studies.

\begin{figure}
    \centering
    \includegraphics[trim=0in 0in 0.5in 0in,clip,width=\textwidth]{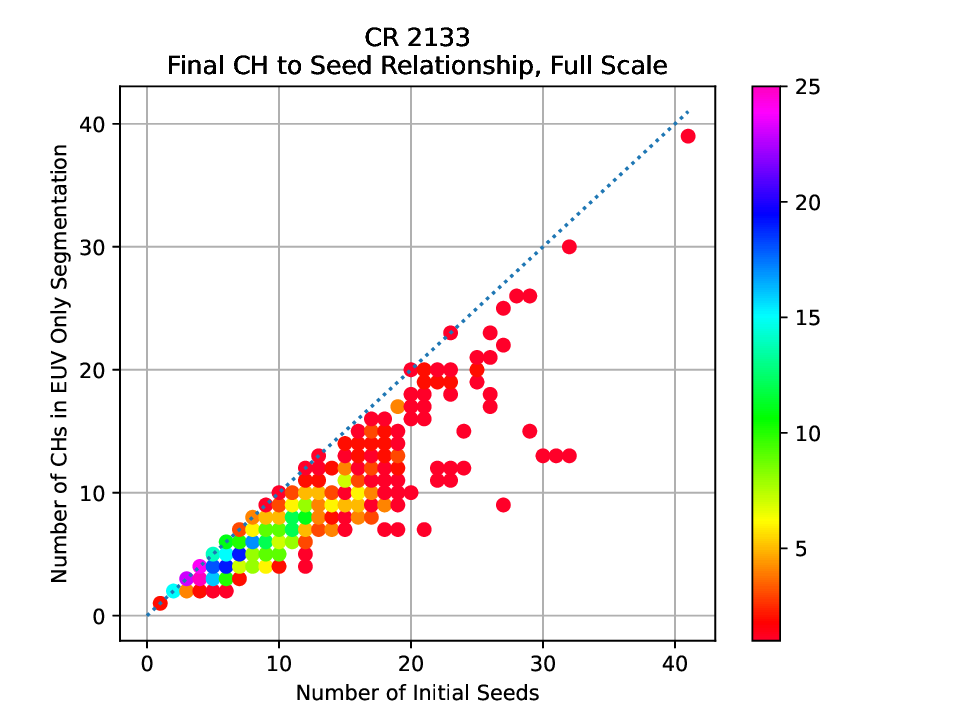}
    \caption{Scatter plot showing the number of regions in the initial seed (x-axis) and in the final segmentation (y-axis) for EUV-only (ACWE) segmentation at full scale for CR 2133. The color of each point denotes the number of images with that specific number of input and output regions. The diagonal line is the $x=y$ line, representing the case where there is one region in the final segmentation for every region in the initial seed.}
    \label{fig:SeedSegCountsFull}
\end{figure}

\begin{figure}
    \centering
    \includegraphics[trim=0in 0in 0.5in 0in,clip,width=\textwidth]{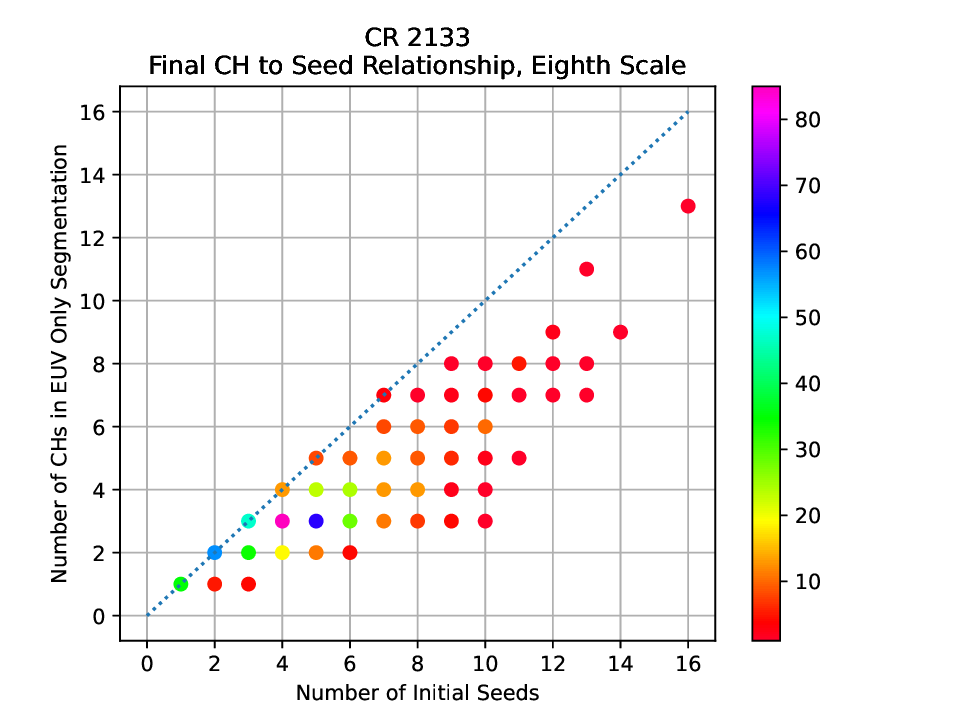}
    \caption{Scatter plot showing the number of regions in the initial seed (x-axis) and in the final segmentation (y-axis) for EUV-only (ACWE) segmentation at one-eighth scale for CR 2133. The color of each point denotes the number of images with that specific number of input and output regions. The diagonal line is the $x=y$ line, representing the case where there is one region in the final segmentation for every region in the initial seed.}
    \label{fig:SeedSegCountsEight}
\end{figure}

First, regardless of scale, multiple regions within the initial seed often contribute to a single CH or filament. The effects of this are seen in Figure~\ref{fig:MagSamplesCR2133}, where the use of magnetogram data minimized evolution of filament regions, resulting in small areas around the initial seeds rather than the unified regions seen in ACWE segmentations. Further demonstration of this behavior is shown by counting the number of regions in the seed and comparing that to the number of regions in the final ACWE segmentation. This comparison for CR 2133 is provided for full-scale (Figure~\ref{fig:SeedSegCountsFull}) and one-eighth-scale (Figure~\ref{fig:SeedSegCountsEight}) ACWE segmentations. In both figures the diagonal line represents the case where one region (CH or filament) is produced in the final segmentation for each region in the seed. Any case above this line indicates more regions in the final segmentation (region splitting), while any case below the line indicates more regions in the initial seed compared to the final contour (region merging). We note that region splitting does not occur, however region merging is prevalent.

\begin{figure}
    \centering
    \includegraphics[trim=0in 0in 0.5in 0in,clip,width=\textwidth]{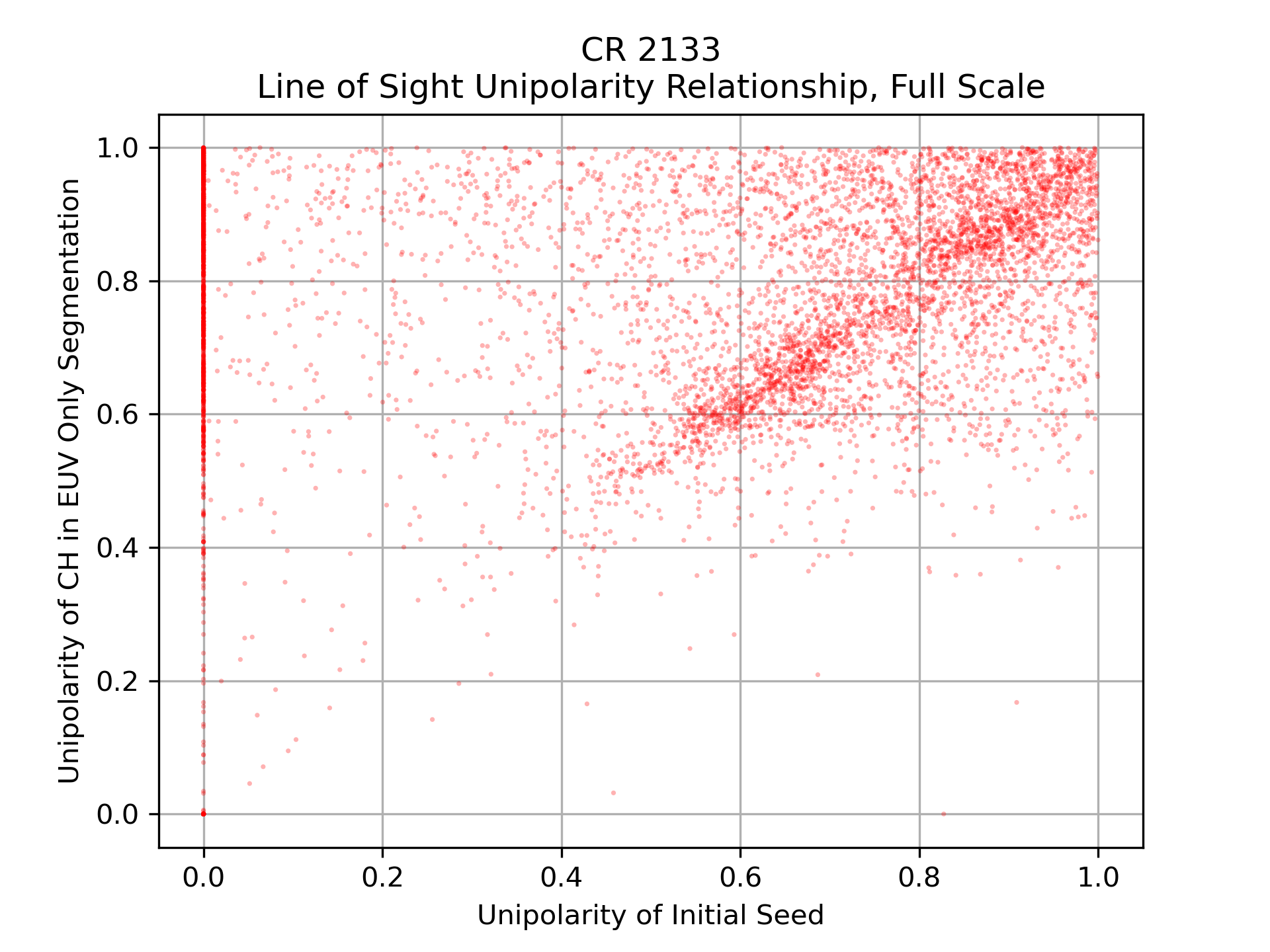}
    \caption{Scatter plot showing the LOS unipolarity of seeds (x-axis) and corresponding regions in the final segmentation (y-axis) for EUV-only (ACWE) segmentation at full scale for all seed regions in CR 2133.}
    \label{fig:LoSUniFull}
\end{figure}

\begin{figure}
    \centering
    \includegraphics[trim=0in 0in 0.5in 0in,clip,width=\textwidth]{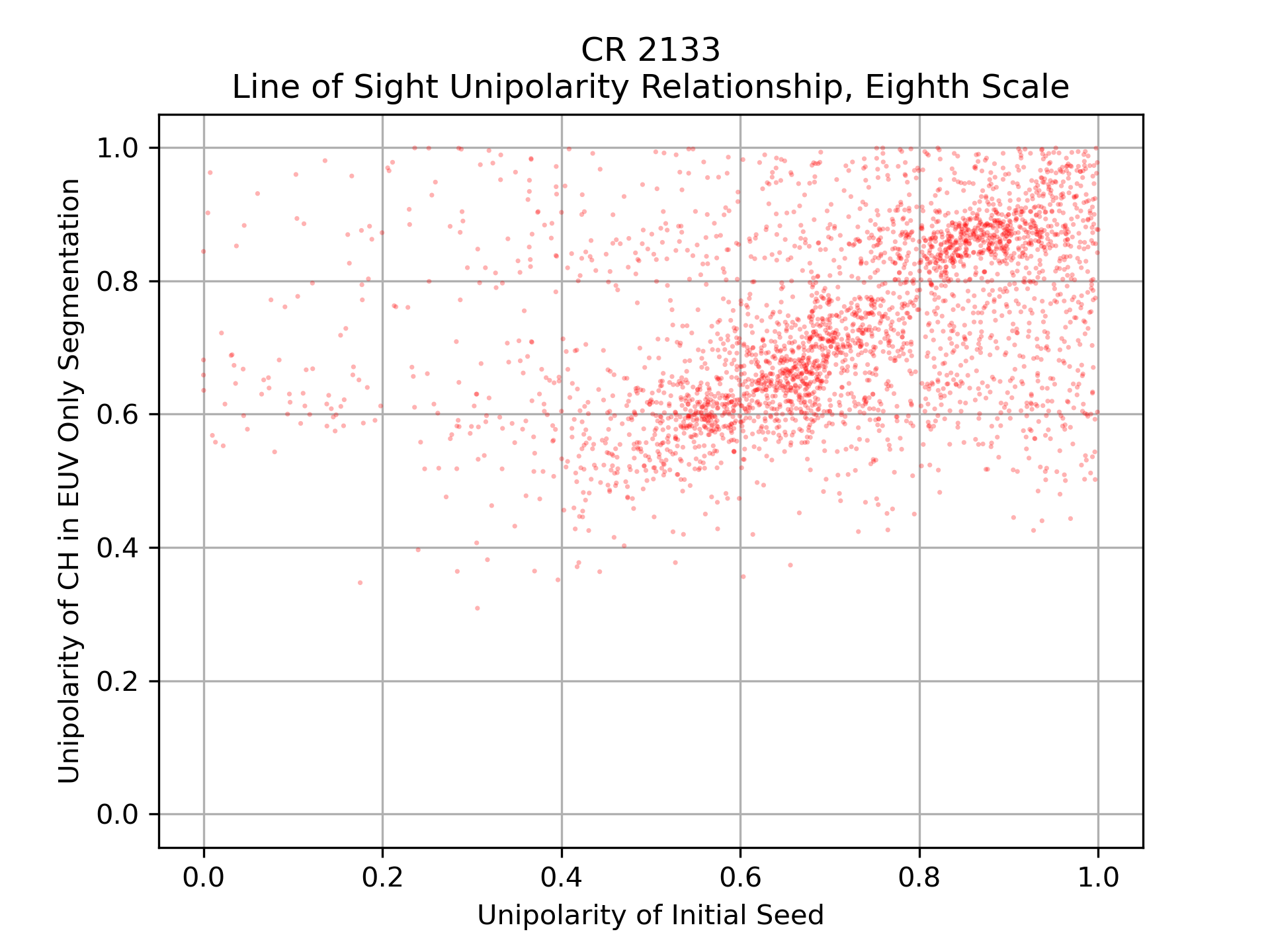}
    \caption{Scatter plot showing the LOS unipolarity of seeds (x-axis) and corresponding regions in the final segmentation (y-axis) for EUV-only (ACWE) segmentation at one-eighth-scale for all seed regions in CR 2133.}
    \label{fig:LoSUniEight}
\end{figure}

Multiple seeds contributing to a single CH is only problematic in that it exacerbates the second issue: small-scale statistics. It is reasonable to assume that regions in the initial contour with close proximity are part of the same final region, however, we note that regions with a large distance between them can belong to the same solar feature. In the counts for Figures~\ref{fig:SeedSegCountsFull} and~\ref{fig:SeedSegCountsEight}, we account for proximity as we did in \cite{grajeda2023}. This is done by defining a region as all pixels that will become a single eight-connected component when dilated with a $40\times40$~pixel square at full scale, or a $5\times5$~pixel square at one-eighth scale. This same definition is applied in Figure~\ref{fig:LoSUniFull} (full scale) and Figure~\ref{fig:LoSUniEight} (one-eighth scale) which show scatter plots of initial versus final unipolarity for each region in the ACWE segmentations.

We note that both Figures~\ref{fig:LoSUniFull} and~\ref{fig:LoSUniEight} indicate regions with a high final (y-axis) value across all initial (x-axis) unipolarity values. This indicates that bipolar regions may appear to have any unipolarity, including high unipolarity, when evaluating the seed. We also note, referring to the cluster along the y-axis of Figure~\ref{fig:LoSUniFull}, that regions reporting high unipolarity in the seed are especially prevalent at the full-scale resolution. This explains why the initial seeds in filament regions remain, even when evolving using both EUV and magnetogram data in QUACK. The significantly smaller cluster on the y-axis of Figure~\ref{fig:LoSUniEight} provides further indication that seeding at one-eighth-scale resolution helps minimize filament contamination. It should also be stated that the results in both Figures~\ref{fig:LoSUniFull} and~\ref{fig:LoSUniEight} contain all seeds across all 618 observations in CR~2133. For this reason, the same CH region is represented multiple times: once for every seed in every observation, as observed from one hour to the next.  The spread, specifically along the x-axis of both figures, further indicates that subtle changes to the seed can strongly affect the apparent unipolarity of a region when evaluating the seed.  These subtle changes to the seed are due to a handful of pixels being included or excluded due to minute variations in EUV intensity from one hour to the next and exacerbate the issue of inaccurate initial polarity.

In summary, multiple seeds may contribute to a final evolved region (CH or filament). Furthermore, seeds may provide an inaccurate estimate of the unipolarity of the final region, mainly due to small-sample statistics.  In light of these results, we consider a means to remove regions associated with filaments.

\subsection{Determining Region Evolution}
\label{sec:FindingFiliments}

To determine if filaments can be removed during evolution, instead of as a post-hoc process, the unipolarity of every region as a function of number of iterations was calculated for both ACWE and QUACK segmentations. This was achieved by seeding and iterating at one-eighth-scale, saving a copy of the contour after each iteration, resizing the output to full-scale resolution, and calculating the LOS unipolarity of each region using the original, full-scale HMI magnetogram.

\begin{figure}
    \centering
    \subfloat[Example of quickly changing unipolarity]{\includegraphics[trim=1.25in 0in 1.5in 0in,clip,width=\textwidth]{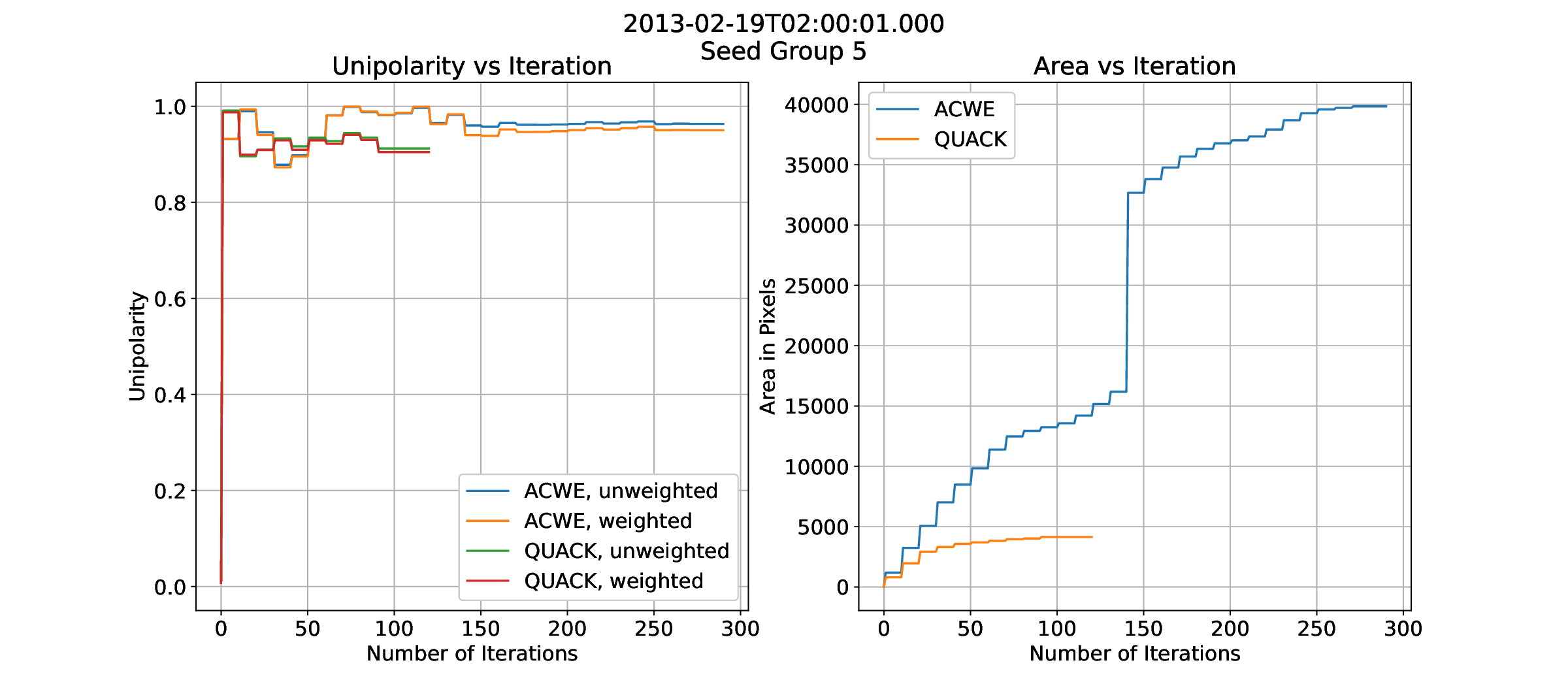}\label{fig:eva}}\\
    \subfloat[Example of slowly changing unipolarity]{\includegraphics[trim=1.25in 0in 1.5in 0in,clip,width=\textwidth]{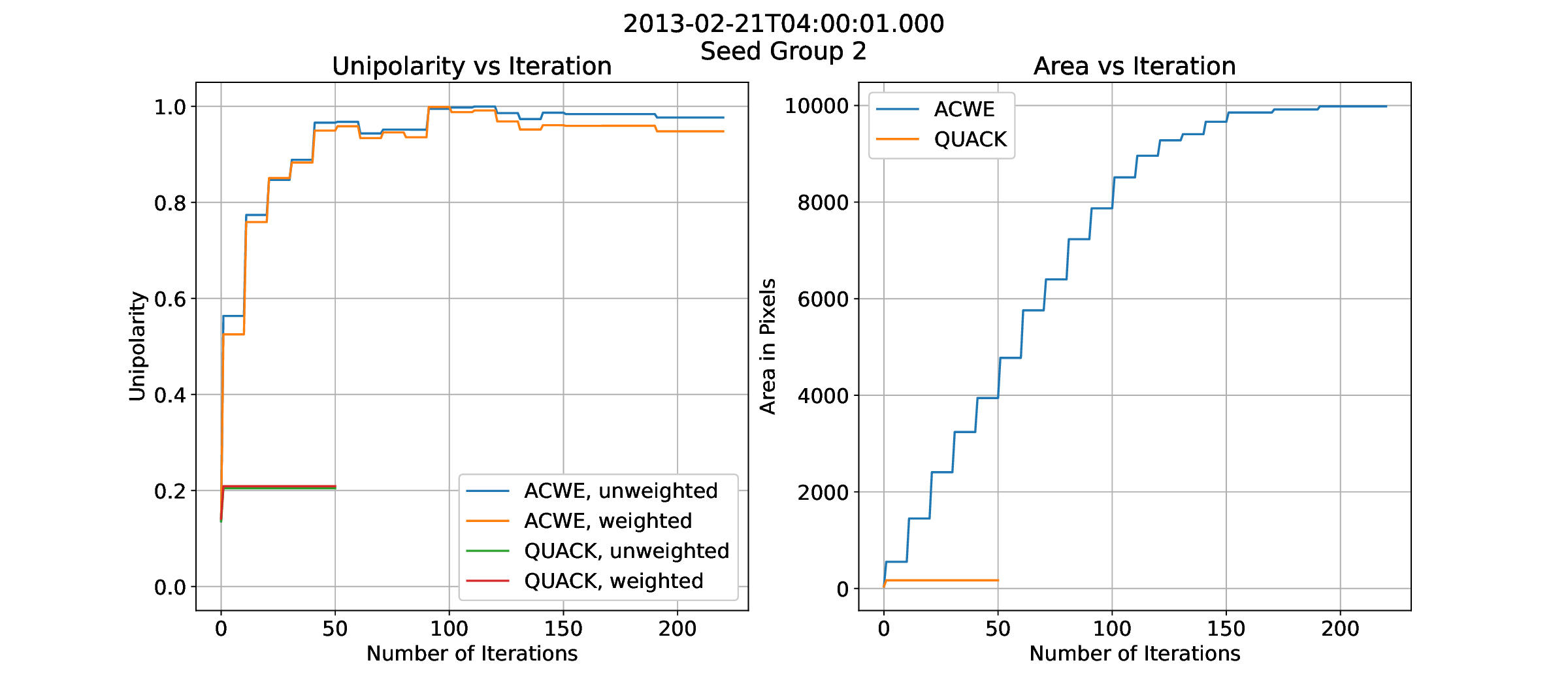}\label{fig:evb}}
    \caption{Region unipolarity (left) and region size in pixels (right) as a function of iteration for ACWE and QUACK segmentations generated at one-eighth scale resolution.}
    \label{fig:Evolution}
\end{figure}

Figure~\ref{fig:Evolution} provides two examples of the change in measured unipolarity and change in area (measured in pixels at full-scale resolution) for a filament region. In the left panel of each example, we provide the LOS unipolarity, calculated via Equation~\ref{eq:uni}, and a weighted unipolarity, weighting regions near disk center (where estimates of magnetic field strength are more accurate) using the method outlined in the SunPy documentation \citep{Weights}. In the right panels, we provide the area.

The examples in Figure~\ref{fig:Evolution} were chosen to demonstrate the two primary behaviors noted among filaments in CR 2133. The first case, seen in Figure~\ref{fig:eva}, consists of regions that quickly become bipolar as they evolve. We note that, when this occurs, the spatially-decimated magnetogram further constrains evolution as compared to an EUV-only evolution, preventing further growth of the region. In these cases the filament region can be identified and removed after very few iterations. This case is representative of most filaments in CR 2133.

The second case, seen in Figure~\ref{fig:evb}, consists of regions that take several iterations before they can be correctly identified as filaments. In these cases the spatially-decimated magnetogram data often prevents the region from evolving before it can be correctly identified as bipolar. Using EUV-only evolution, these cases typically evolve enough to correctly identify in $<50$ iterations. Due to the presence of filaments with the behavior seen in Figure~\ref{fig:evb}, we remove potential filament regions after evolving the contour for 50 EUV-only iterations. To allow for the variation in measured unipolarity seen in both examples in Figure~\ref{fig:Evolution} and in other similar cases, we remove regions with a unipolarity $U\geq0.8$.

Interestingly, we note in Figure~\ref{fig:Evolution} that the incorporation of magnetic field appears to result in faster convergence.  While definitive reasons for this are unclear, we hypothesize that it is due to two factors.  First, estimates of CHs are now more conservative since the contour is guided away from QS by the unipolarity constraint.  Second, there are now two forces (homogeneity and unipolarity) acting on the contour, guiding it to a stable (albeit smaller) solution faster.

In summary, inclusion of HMI data and the unipolarity term in the energy functional constrains evolution of filament regions to remain small and isolated around the seed (which is determined based solely on EUV intensity).  That constraint, however, can result in too small an area for accurate statistics in the computation of unipolarity, resulting in misidentification of filaments as small CHs.  Thus, a seed filtering process to remove filaments can utilize initial evolution with only EUV data, removal of regions with insufficient unipolarity, and evolution of remaining regions to convergence using EUV and HMI data.

\section{Updated ACWE Segmentation Method}
\label{sec:pipelineSpecific}

\subsection{Updated Pipeline}
Results in Section~\ref{sec:FindingFiliments} demonstrate that the seed does not provide sufficient information to correctly identify filaments. Results in Section~\ref{sec:QUACK_eight} demonstrate that decimated magnetograms still constrain CH evolution to highly unipolar regions.  Based on these results, we develop the following QUACK pipeline:
\begin{enumerate}
    \item Open and, if needed, update all observations to ensure correct header information, consistent solar radii, and that all observations are oriented so that solar north is at the top of the image.
    \item Resize all observations to $512\times512$~pixels.
    \item Correct EUV observation(s) for limb brightening.
    \item Generate initial seed using EUV observation(s).
    \item Combine all observations into a vector-valued image.
    \item Mask off-disk areas and evolve using only the EUV observation(s) until the the unipolarity of each region can be accurately estimated, approximately 50 iterations.  These initial iterations set $\lambda^{\rm{U}+}_j=\lambda^{\rm{U}-}_j=0$ where $j$ is the index of the magnetogram data, and evolve using Equation~\ref{eq:acwevu} as the energy functional.
    \item Identify and remove bipolar regions.
    \item Continue evolving using both EUV and magnetogram data (by setting $\lambda^{\rm{U}+}_j$ and $\lambda^{\rm{U}-}_j$ to non-zero values) until convergence.
\end{enumerate}
We note that every ten iterations, both when evolving only EUV data and when evolving with both EUV and magnetogram data, we reset the intensity of the off-disk regions to the channel-specific mean of the non-CH, on-disk region.

\subsection{Results}
In this section, unless otherwise stated, we use the EUV 193~{\AA} observation, seed using $\alpha=0.3$, and perform EUV-only evolution for $50$ iterations using the homogeneity parameters from Section~\ref{sec:QUACK_full} before removing regions with a LOS unipolarity $U\geq0.8$. We perform the full EUV+HMI magnetogram evolution on the remaining regions using the weights from Section~\ref{sec:QUACK_full}.

\subsubsection{Primary Results}
\begin{figure}
    \centering
    \subfloat[IOU]{\includegraphics[trim=0in 1.2in 0in 1.3in,clip,width=0.45\textwidth]{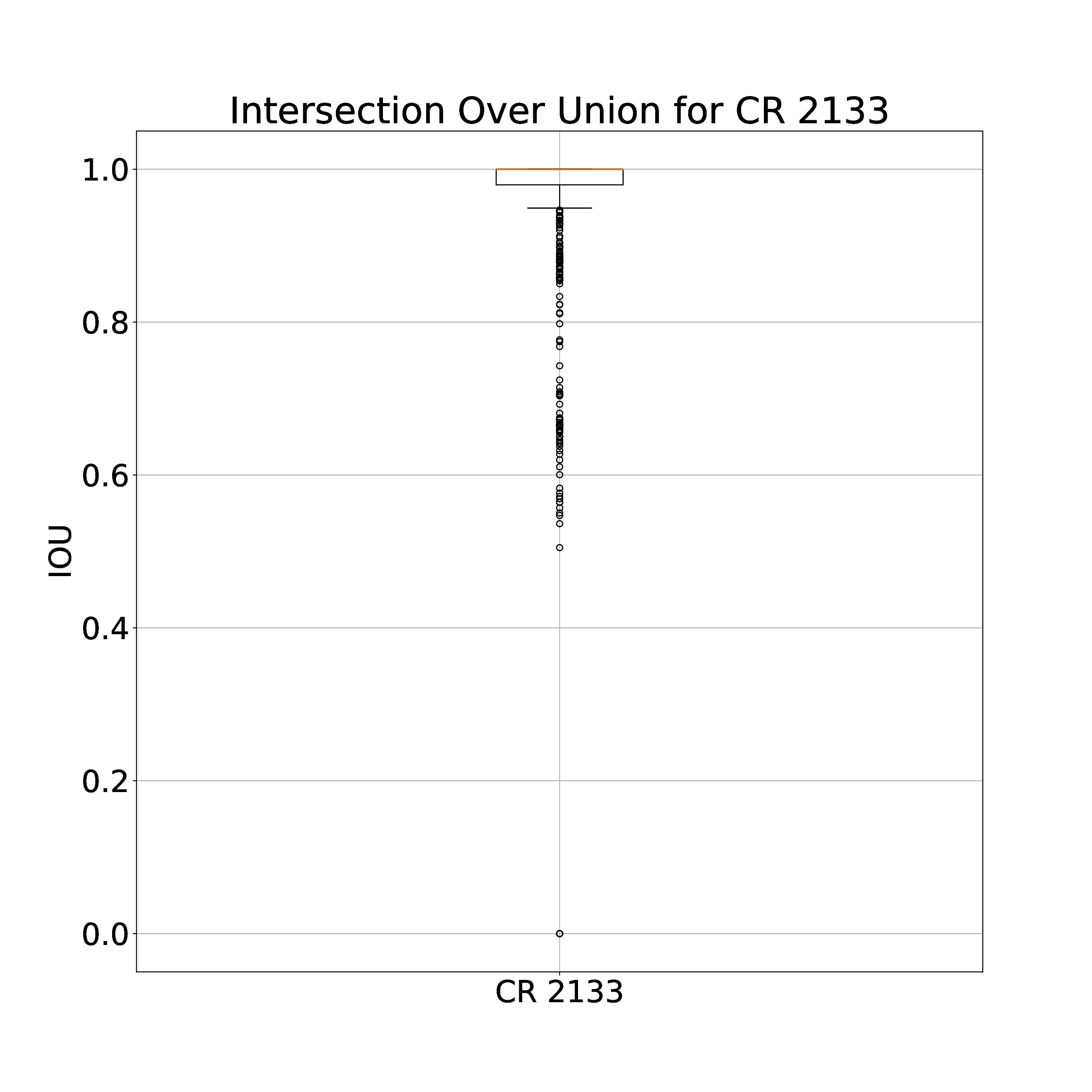}}~
    \subfloat[SSIM]{\includegraphics[trim=0in 1.2in 0in 1.3in,clip,width=0.45\textwidth]{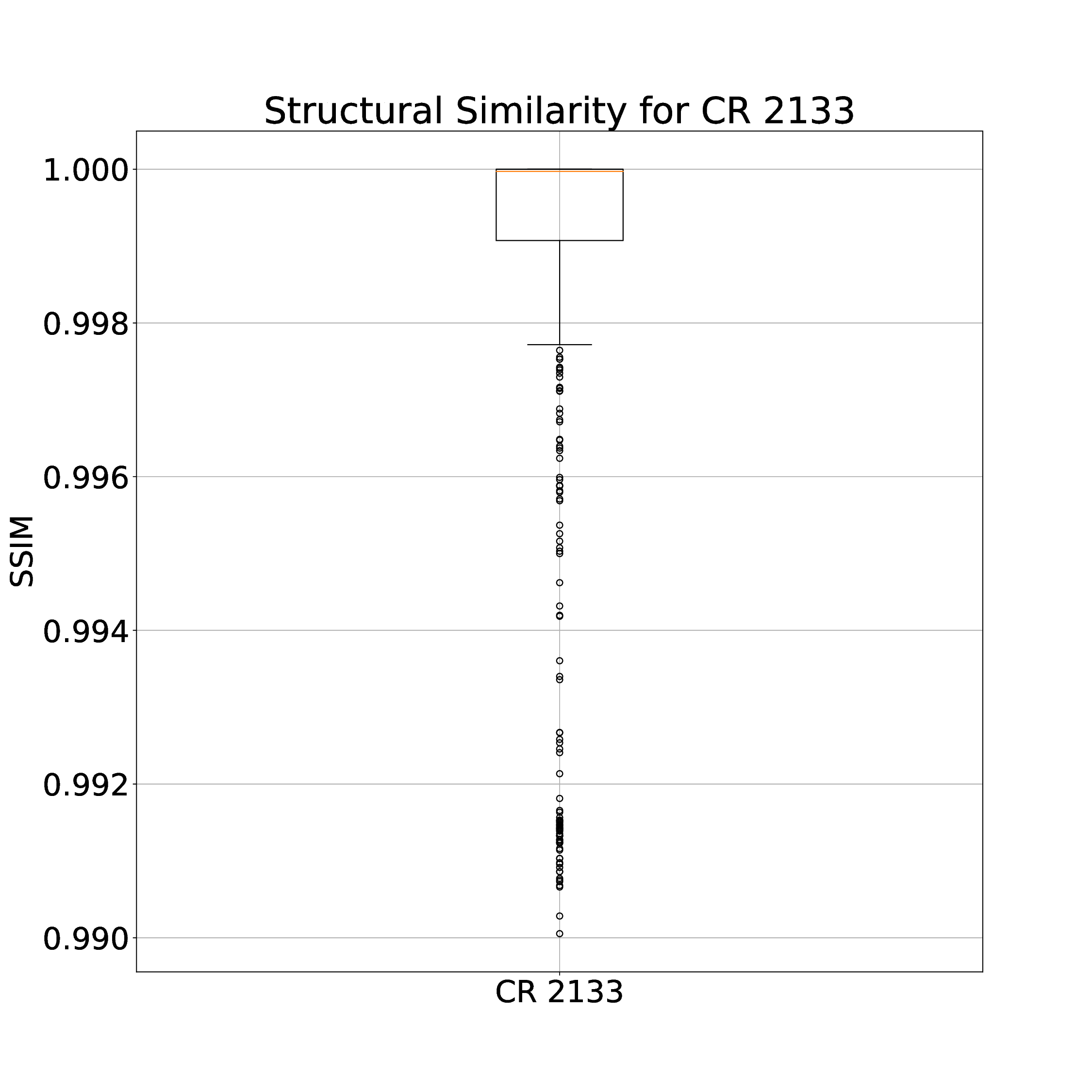}}\\
    \subfloat[GCE]{\includegraphics[trim=0in 1.2in 0in 1.3in,clip,width=0.45\textwidth]{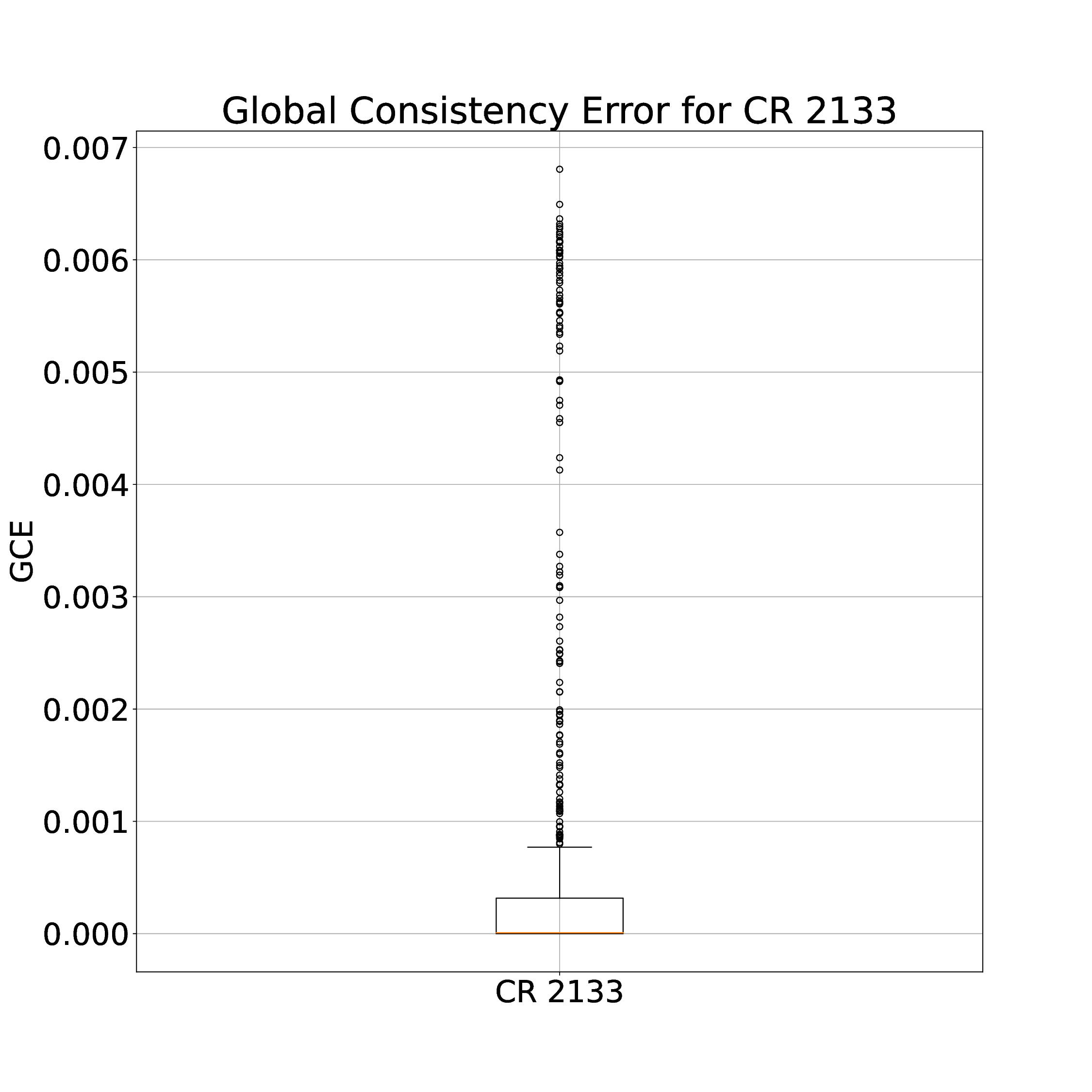}}~
    \subfloat[LCE]{\includegraphics[trim=0in 1.2in 0in 1.3in,clip,width=0.45\textwidth]{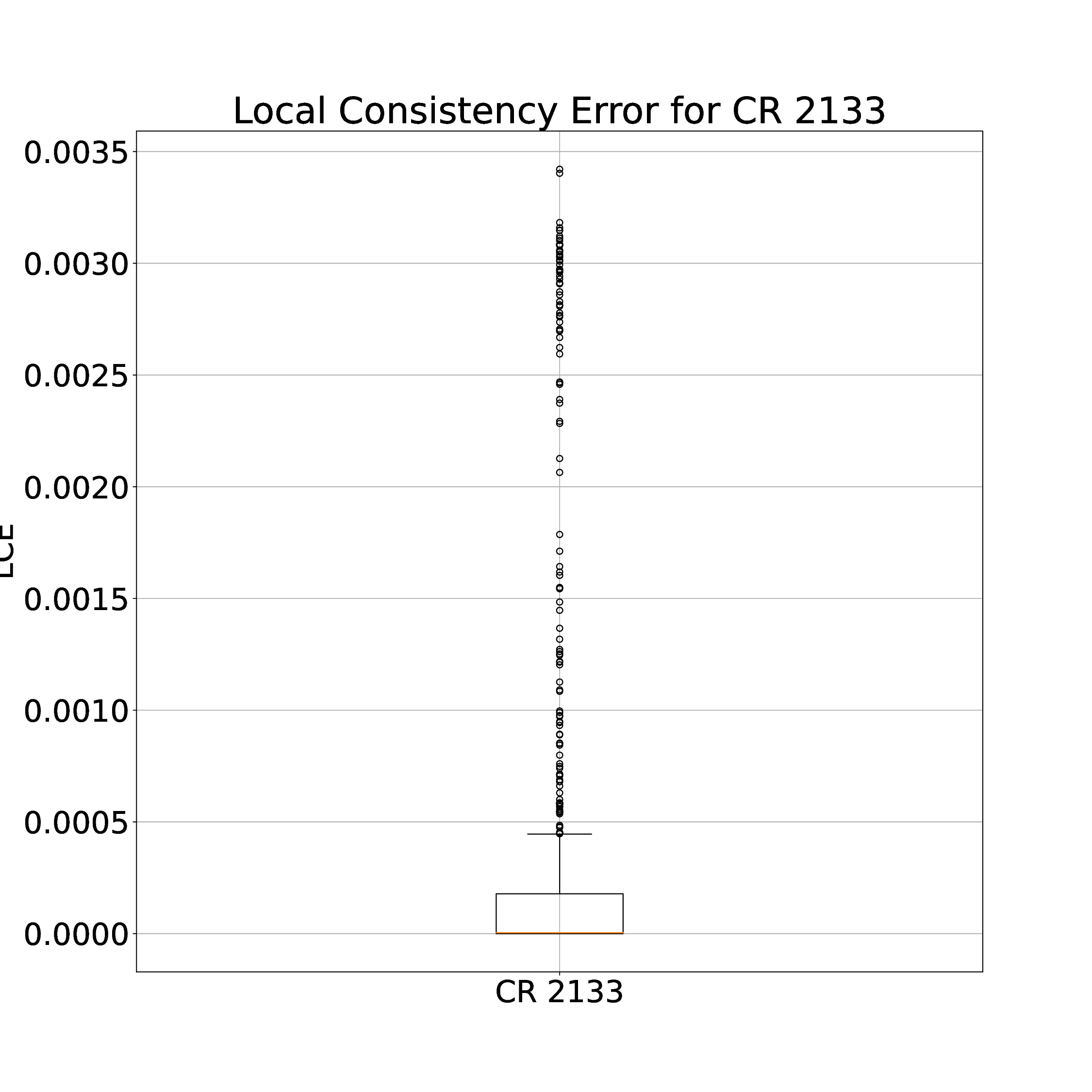}}\\
    \caption{Comparison between one-eighth-scale QUACK segmentations with and without seed filtering for CR 2133. The box outlines the range between Q1 and Q3, with the median value in orange. The whiskers show 1.5 times the interquartile range. Outliers are marked with circles.}
    \label{fig:QUACK_CR2133}
\end{figure}

\begin{figure}
    \centering
    \subfloat[$U_1=0.9553$]{\includegraphics[trim=0in .2in 0in 0in,clip,width=\textwidth]{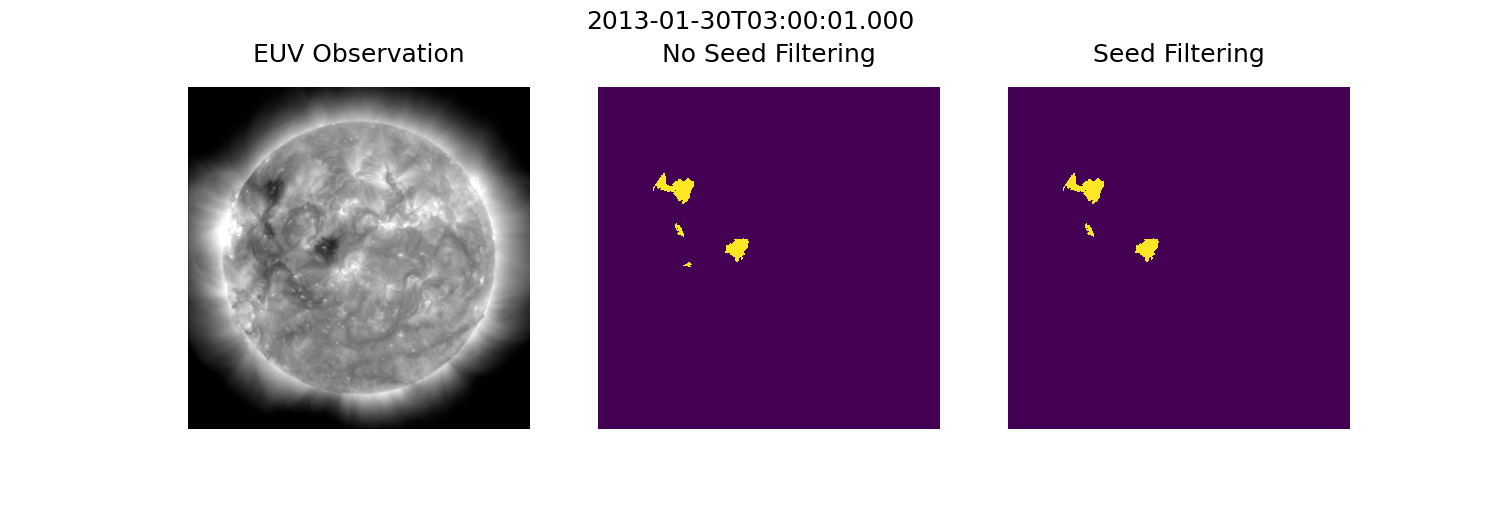}}\\
    \subfloat[$U_1=0.9128$, $U_2=0.9972$]{\includegraphics[trim=0in .2in 0in 0in,clip,width=\textwidth]{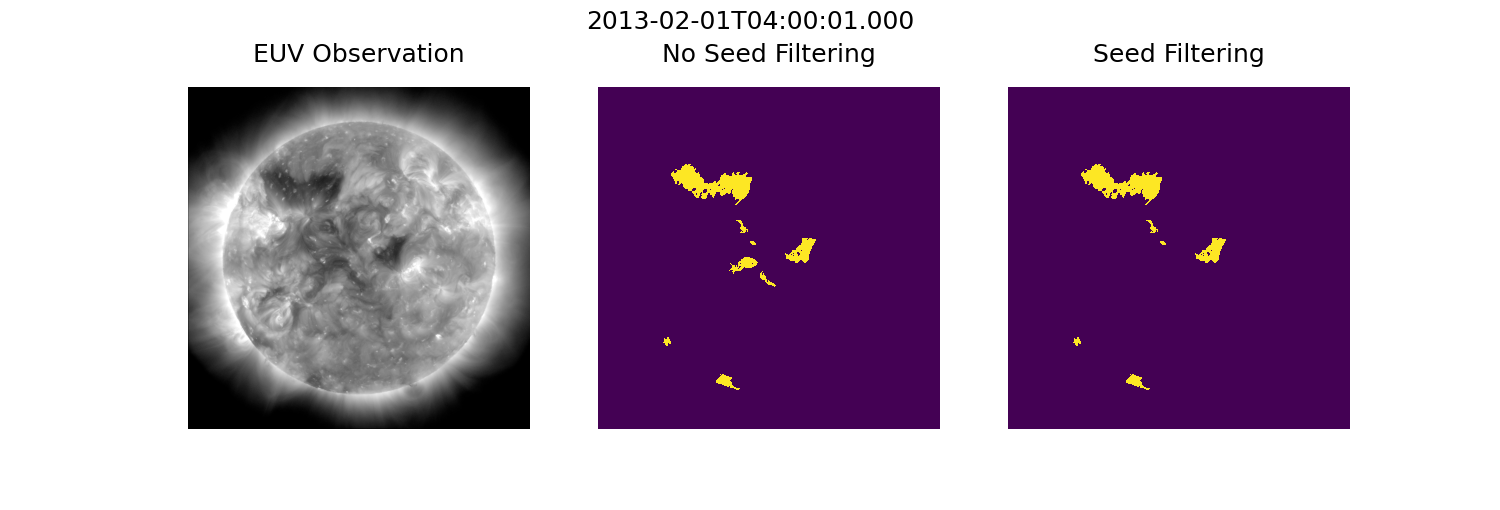}}\\
    \subfloat[$U_1=0.9866$, $U_2=0.8861$]{\includegraphics[trim=0in .2in 0in 0in,clip,width=\textwidth]{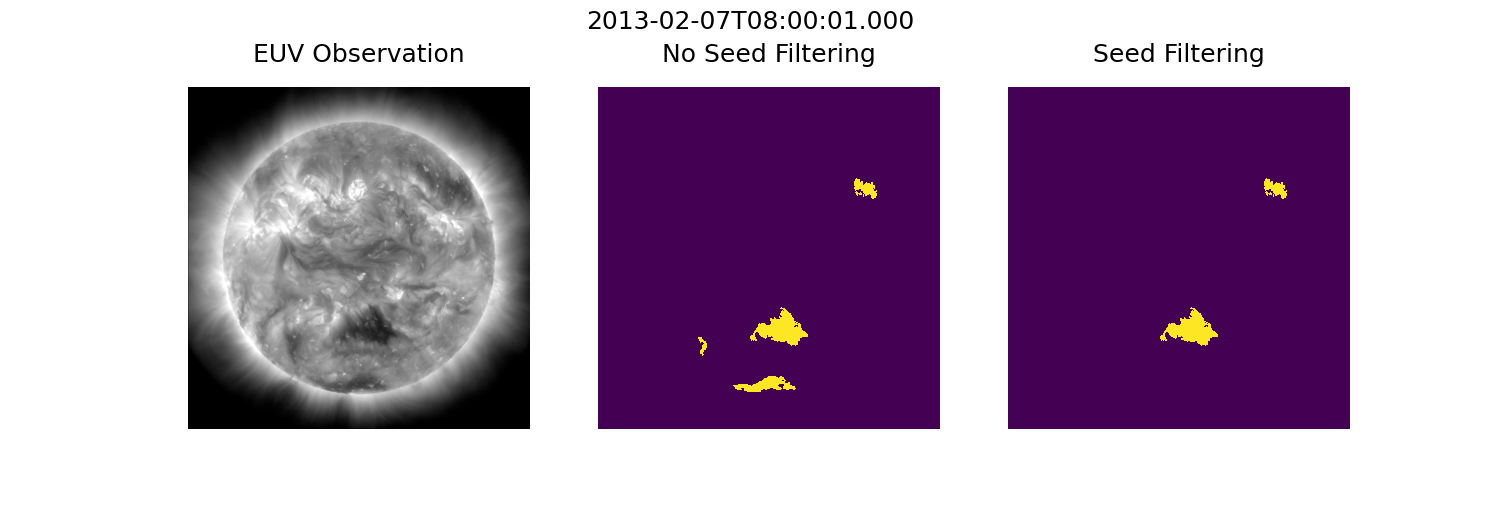}\label{fig:QUACK_samples2133_oops}}
    \caption{Examples of QUACK segmentations from CR 2133 generated without filtering the initial seed (center) and with seed filtering (right). The caption under each example provides the unweighted LOS unipolarity of the missing regions in order of right to left, top to bottom.}
    \label{fig:QUACK_samples2133}
\end{figure}

Figure~\ref{fig:QUACK_CR2133} summarizes the similarity between the one-eighth-scale seeded and evolved QUACK segmentations with and without seed filtering for CR 2133. We note that the median IOU and SSIM are both $1$ and that the median GCE and LCE are both $0$, indicating that the vast majority of segmentations are identical. This is consistent with the expectation that seeding at one-eighth-scale already removed the majority of filaments (see Section~\ref{sec:QUACK_eight}). Figure~\ref{fig:QUACK_samples2133} provides samples of the segmentations with differences between inclusion and exclusion of seed filtering. We note that regions missing as a result of seed filtering are regions that appear highly bipolar based on LOS, full-scale magnetogram data. Referencing Figure~\ref{fig:QUACK_samples2133_oops}, we further note that this can, in rare cases, result in the loss of observations near disk edge. Exploration of a variable unipolarity threshold that is a function of distance from disk center is therefore an area of future research.

\begin{figure}
    \centering
    \includegraphics[trim=1.2in .7in 1.5in 1in,clip,width=\textwidth]{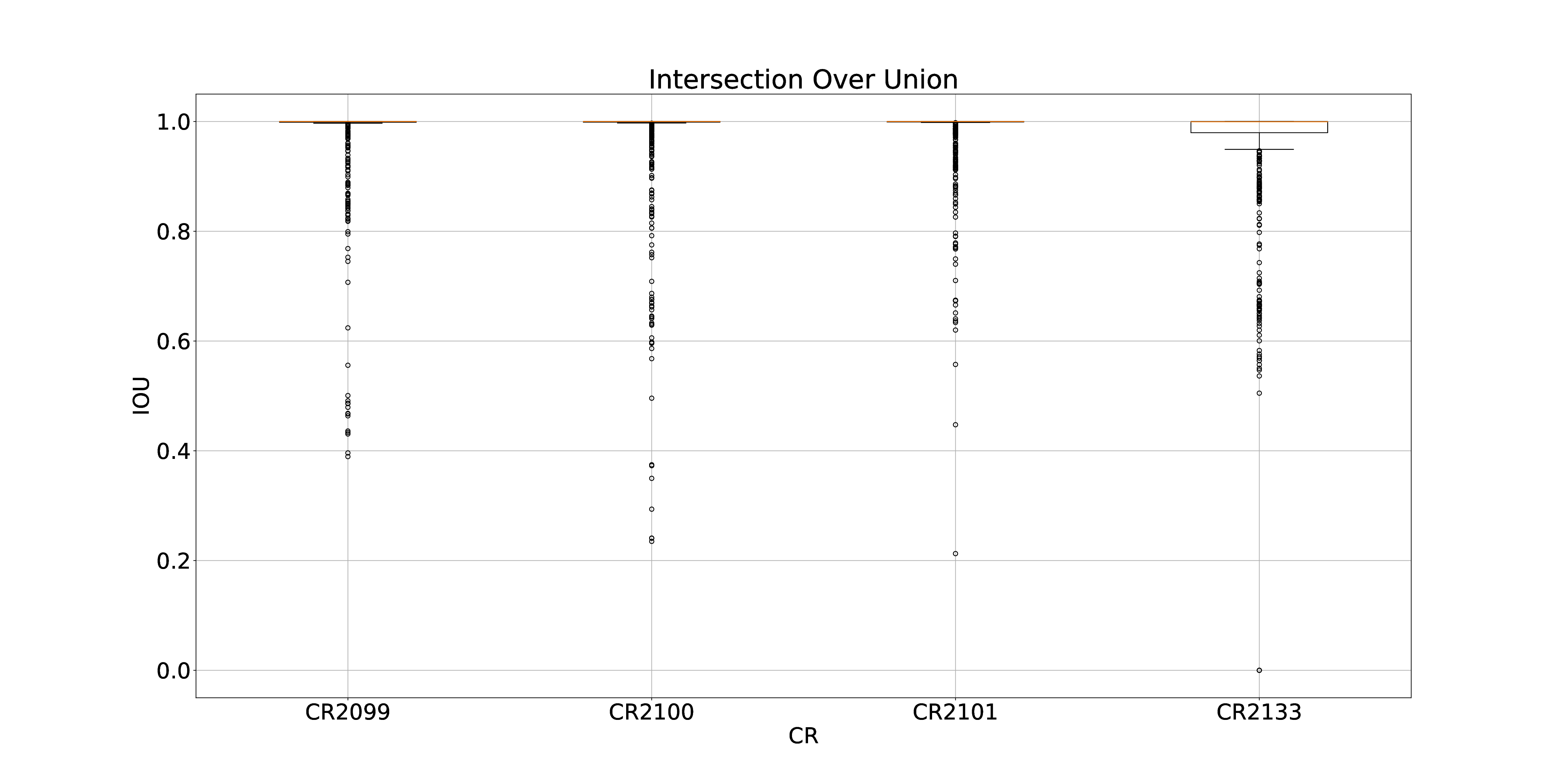}\\
    \caption{IOU between one-eighth-scale QUACK segmentations with and without seed filtering for each CR. The box outlines the range between Q1 and Q3, with the median value in orange. The whiskers show 1.5 times the interquartile range. Outliers are marked with circles.}
    \label{fig:QUACK_IOU_box}
\end{figure}

\begin{figure}
    \centering
    \subfloat[CR 2099]{\includegraphics[trim=0in .7in 0in 1.2in,clip,width=0.45\textwidth]{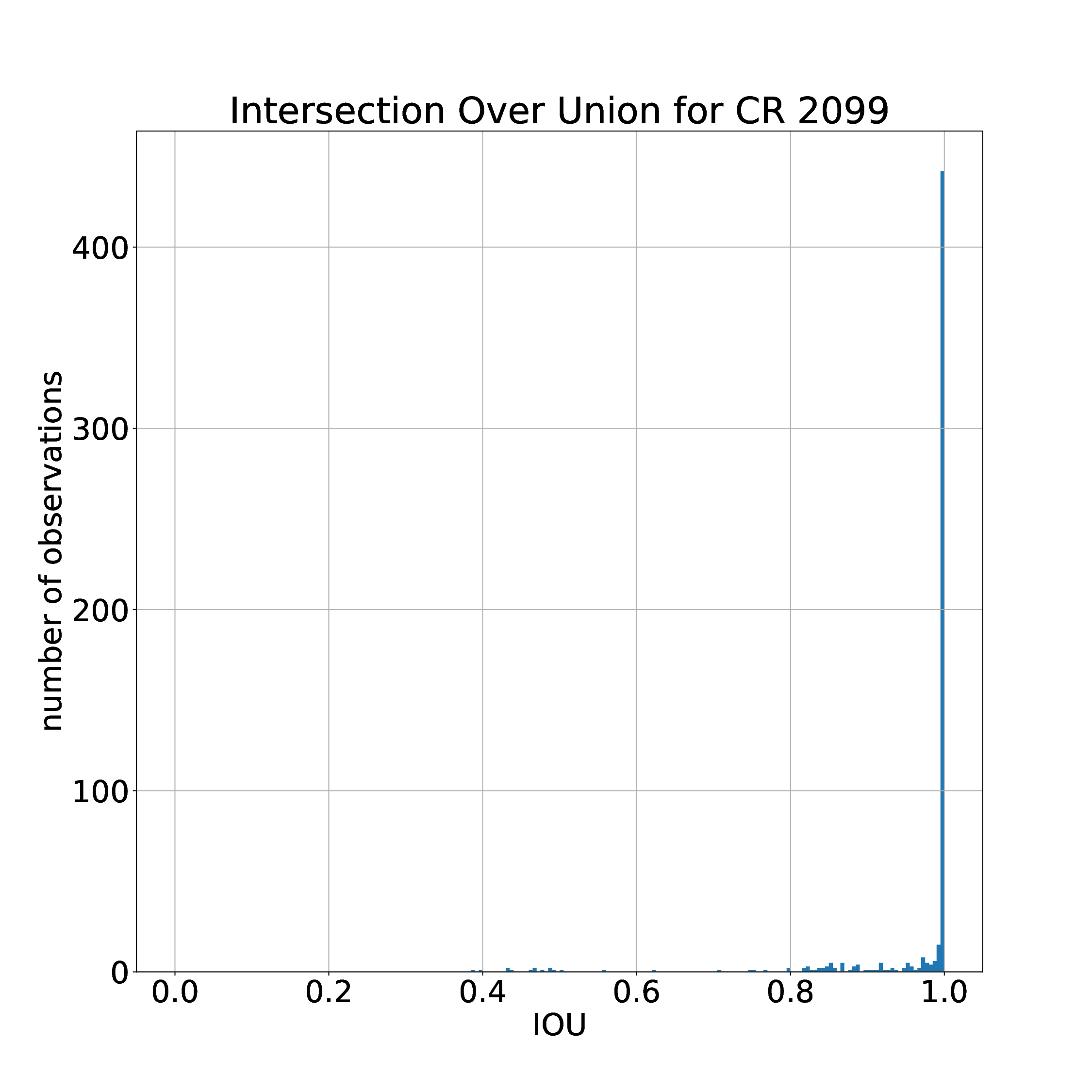}}~
    \subfloat[CR 2100]{\includegraphics[trim=0in .7in 0in 1.2in,clip,width=0.45\textwidth]{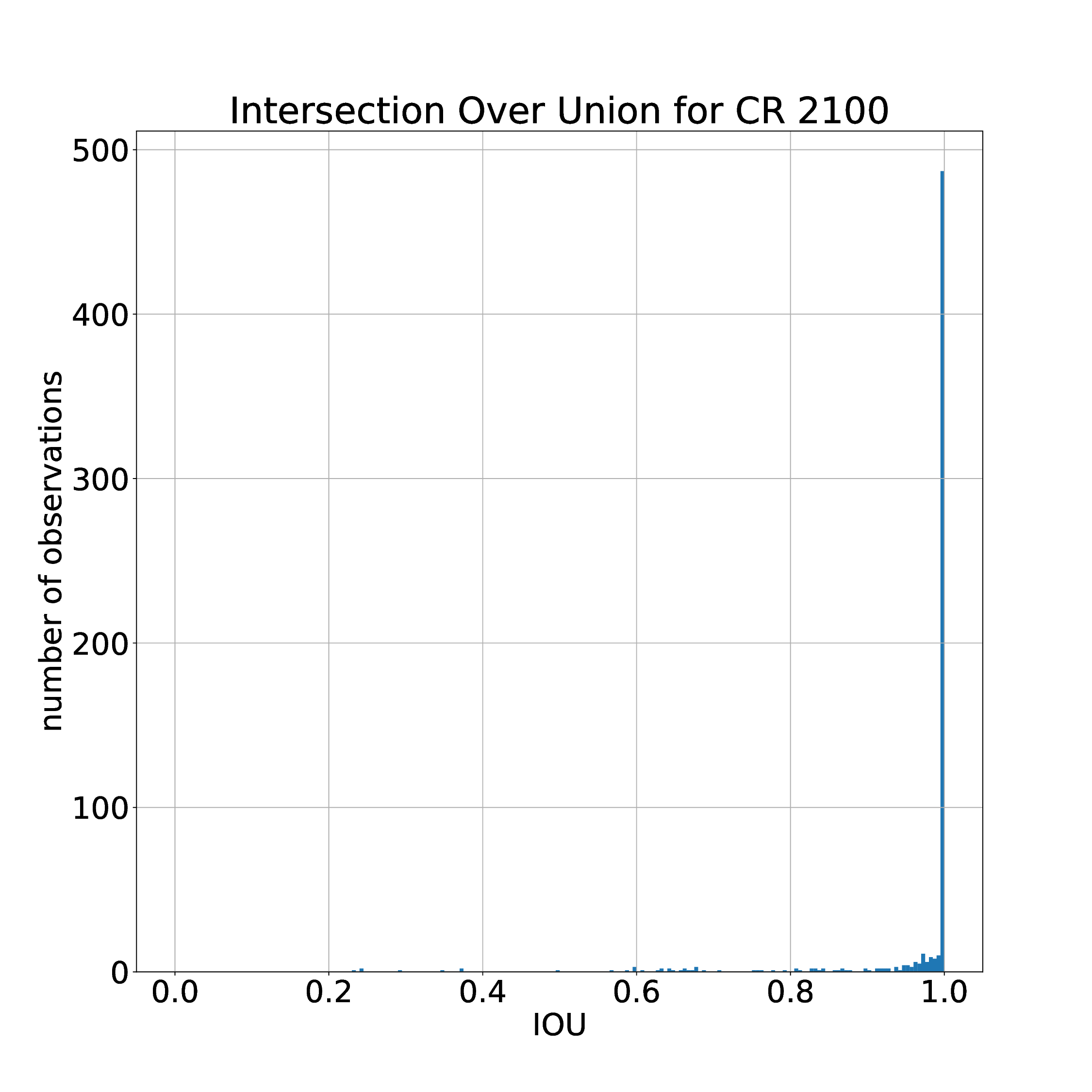}}\\
    \subfloat[CR 2101]{\includegraphics[trim=0in .7in 0in 1.2in,clip,width=0.45\textwidth]{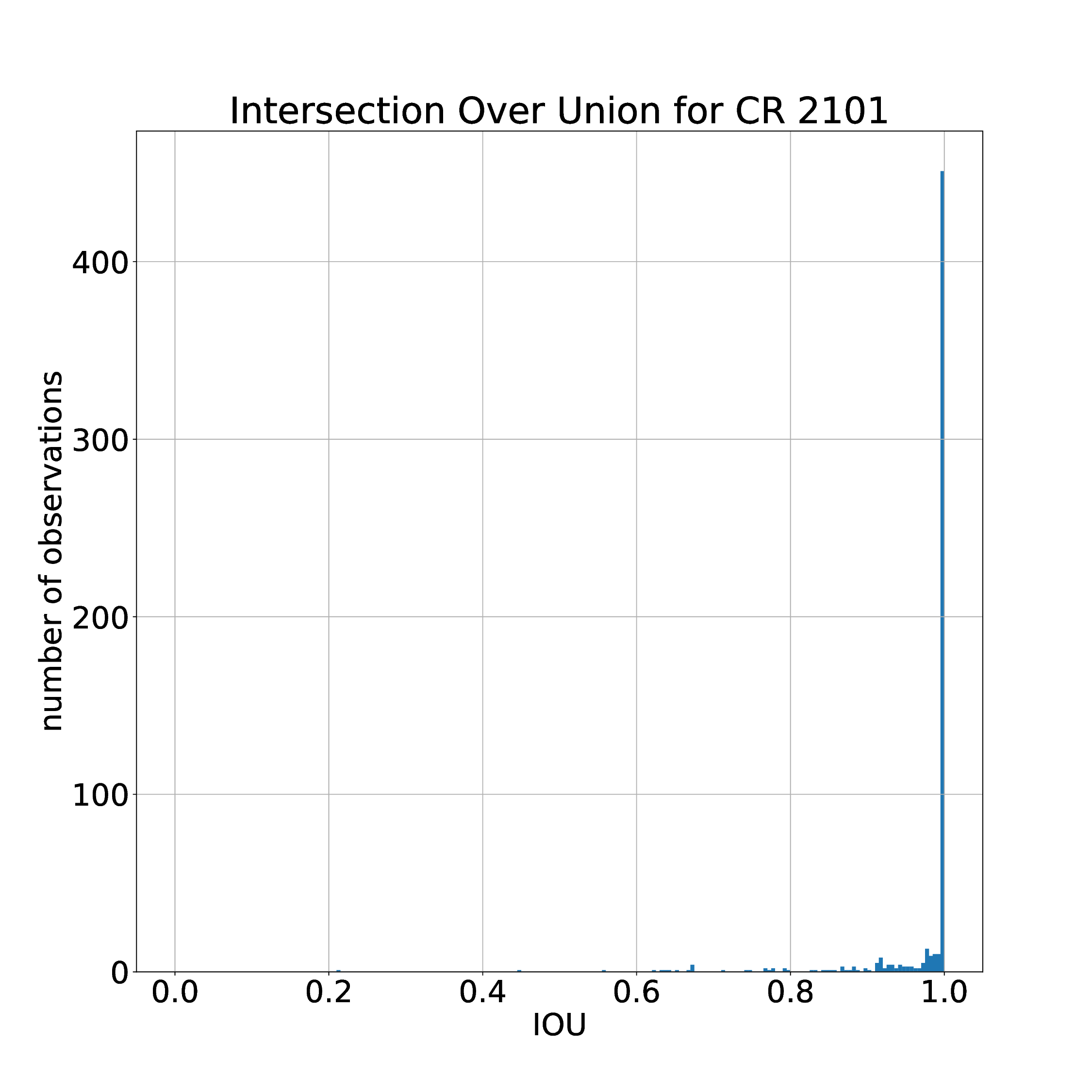}}~
    \subfloat[CR 2133]{\includegraphics[trim=0in .7in 0in 1.2in,clip,width=0.45\textwidth]{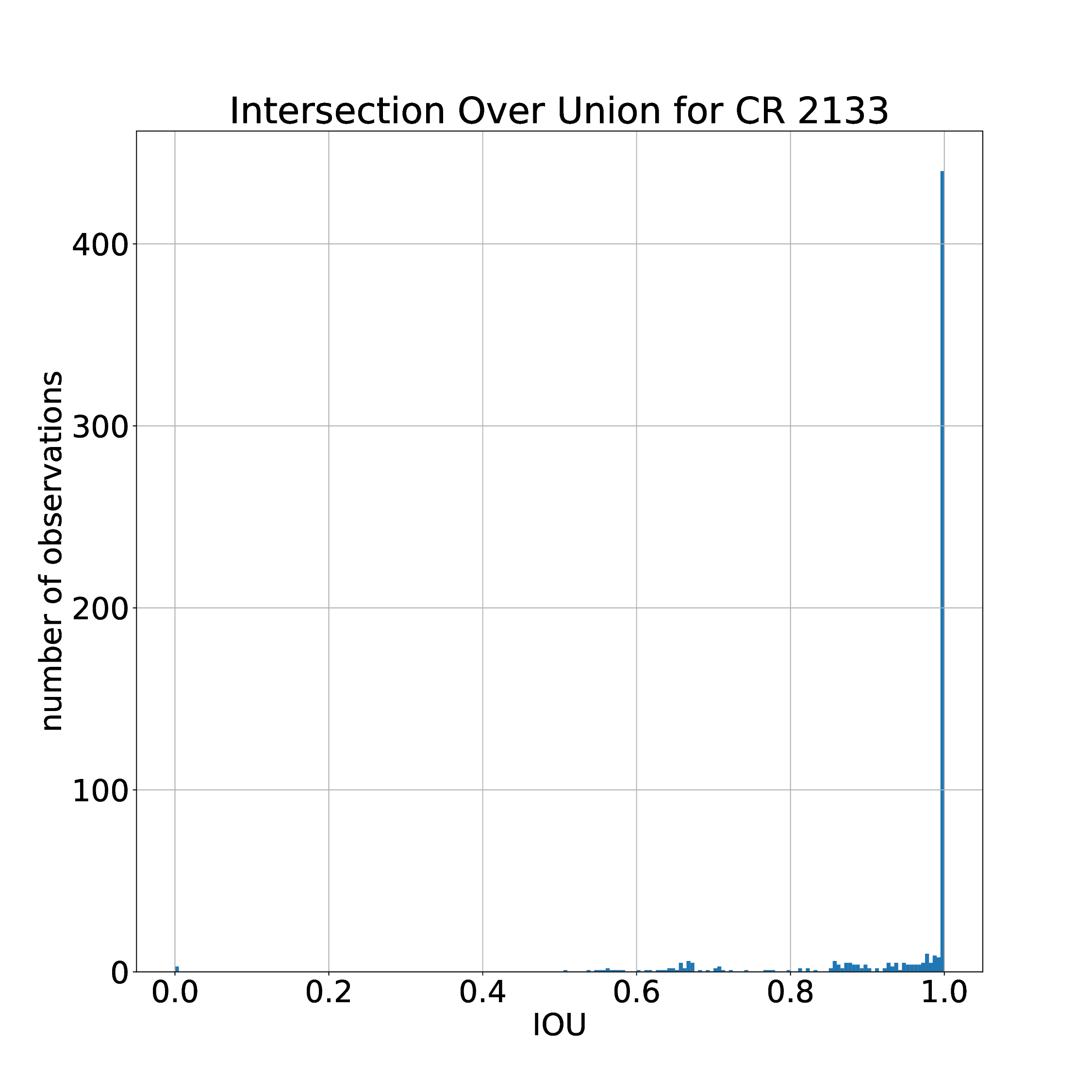}}\\
    \caption{IOU between one-eighth-scale QUACK segmentations with and without seed filtering for each CR.}
    \label{fig:QUACK_IOU_hist}
\end{figure}

Noting that IOU is the most stringent of the four metrics, we provide comparisons between QUACK segmentations with and without seed filtering for all four CRs as box plots in Figure~\ref{fig:QUACK_IOU_box} and as 200-bin histograms in Figure~\ref{fig:QUACK_IOU_hist}. Both figures indicate that all segmentations with discrepancies appear to be outliers. Visual inspection of those segmentations with discrepancies reveal that the majority of regions missing when using seed filtering, but present when omitting seed filtering, consisted of small regions that appear highly bipolar in full-scale LOS magnetogram data. The remaining regions missing when using seed filtering consist of regions along the disk edge. All regions that were retained appear visually identical. We note that the plots in Figures~\ref{fig:QUACK_IOU_box} and~\ref{fig:QUACK_IOU_hist} present data for four CRs; thus, while we analyze hundreds of segmentations per CR, the effective sample size is smaller in that CHs are expected to remain relatively stable in appearance over a single CR.  We consider a larger sample size in Section~\ref{sec:extended_study}.

\subsubsection{Comparison to ACWE}

In order to further compare QUACK to ACWE we utilize the dataset of \cite{reiss2023}. This dataset contains seven AIA Level 1 EUV images and the 45-second LOS HMI magnetograms with the closest observation time, all preprocessed to ensure alignment between the data products. The results in this section thus represent our first-ever segmentations using 45-second LOS HMI magnetograms (rather than 720-second magnetograms).

To provide a fair comparison between methods, segmentation is performed twice, once using a seed generated with $\alpha=0.3$, and once using $\alpha=0.4$, referred to as ``QUACK03'' and ``QUACK04,''  respectively. These two segmentations will be compared to ``ACWE03'' and ``ACWE04'' of \cite{reiss2023}, respectively, which begin with the exact same seed (with one exception). The two versions of ACWE in \cite{reiss2023} were produced because a threshold of $\alpha=0.3$ does not yield any segmentation for the observation on 2 September 2016. Assuming that the full QUACK pipeline can mitigate erroneous detections (e.g., filaments) in subsequent steps, QUACK will increment $\alpha$ until a seed is generated. For this reason, a seed created with $\alpha=0.33$, automatically chosen by the algorithm, is used for this single observation in the QUACK03 segmentations.

\begin{figure}
    \centering
    \subfloat[One filament removed, one filament reduced]
    {\includegraphics[trim=0in .45in 0in 0in,clip,width=\textwidth]{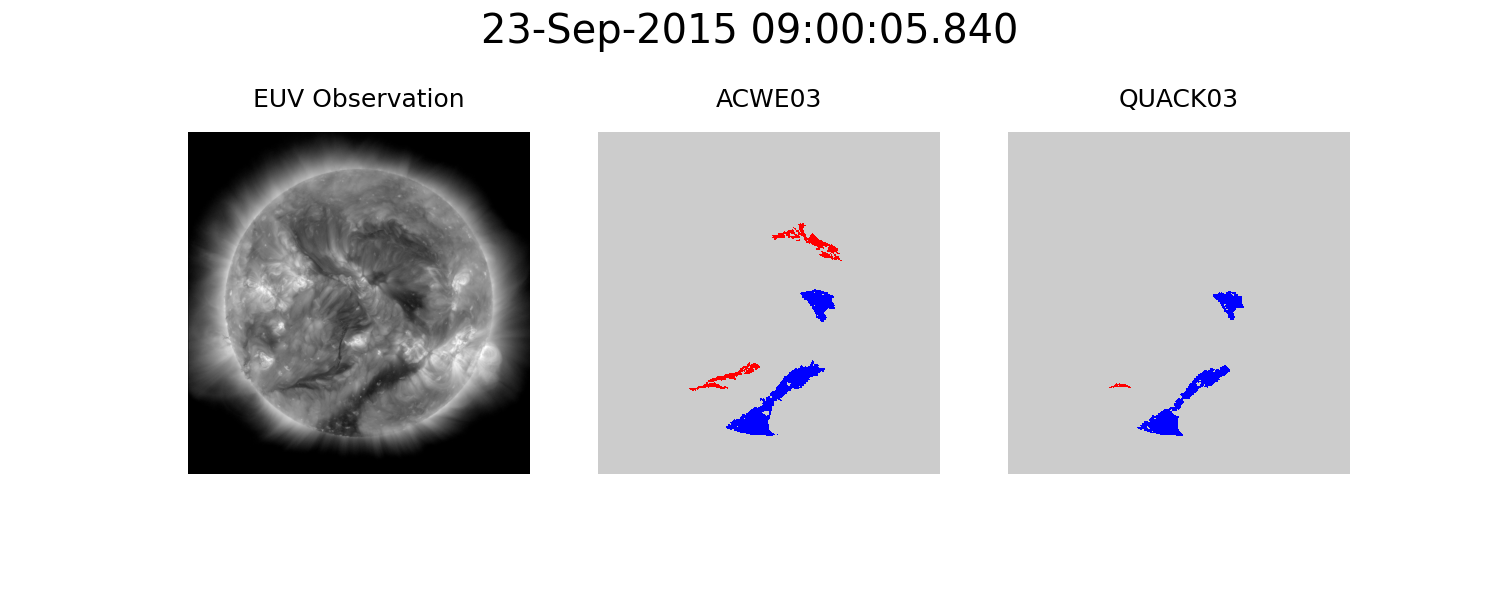}\label{fig:ReissSamplesA}}\\
    \subfloat[One filament reduced]
    {\includegraphics[trim=0in .45in 0in 0in,clip,width=\textwidth]{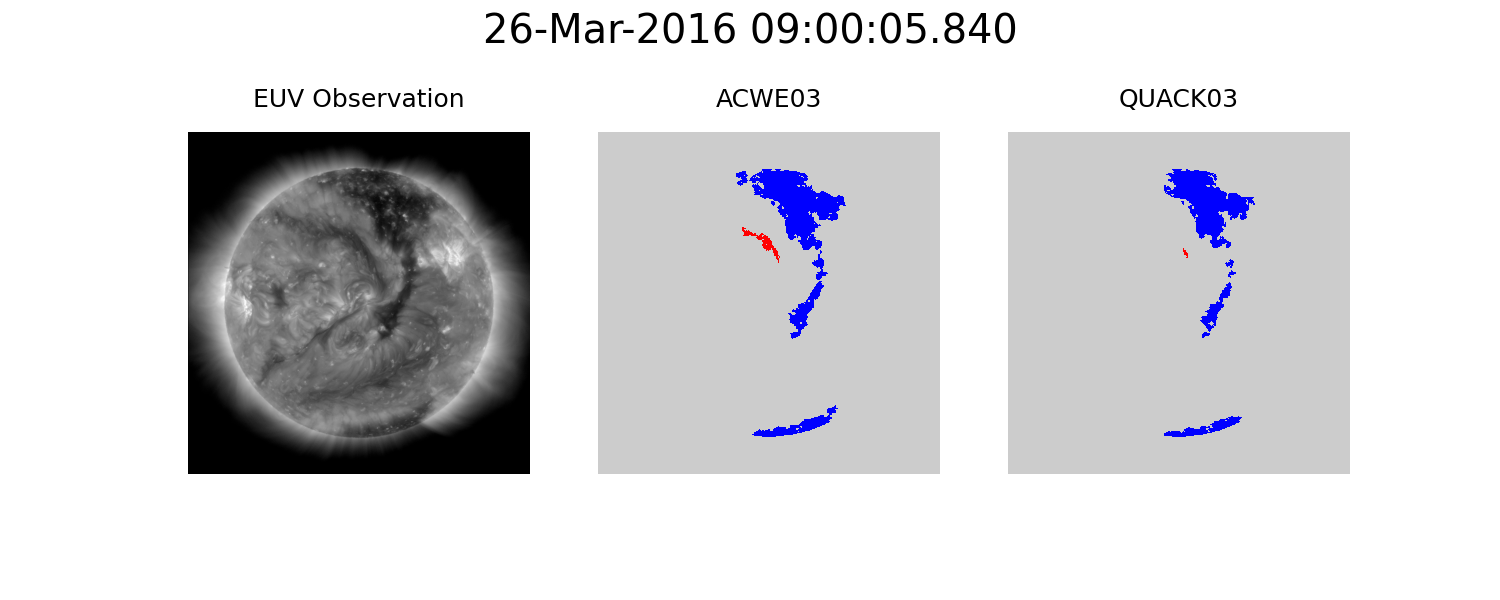}\label{fig:ReissSamplesB}}\\
    \subfloat[Example effect of minimizing growth into QS]
    {\includegraphics[trim=0in .45in 0in 0in,clip,width=\textwidth]{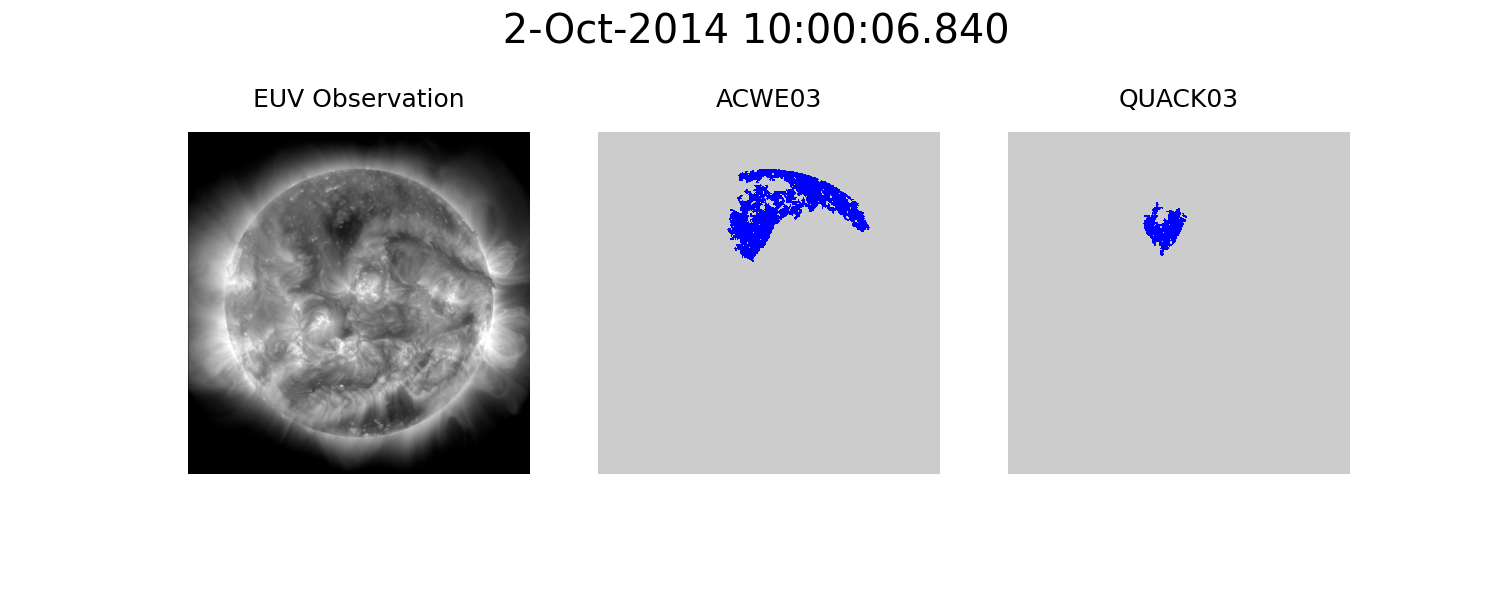}\label{fig:ReissSamplesC}}\\
    \caption{Comparison between ACWE03 and QUACK03 segmentation for cases in the dataset of \cite{reiss2023}. In these examples, regions labeled as CH in~\cite{reiss2023} are blue and regions labeled as filaments in~\cite{reiss2023} are red.}
    \label{fig:ReissSamples}
\end{figure}

Of the 5 filaments present in ACWE03, QUACK03 removed 3 filaments: one each from the observations on 6 June 2015, 11 August 2015, and 23 September 2015. The first retained filament, in the data from 23 September 2015, shown in Figure~\ref{fig:ReissSamplesA}, was reduced from $875$~pixels to $121$~pixels (at $512\times512$~pixel resolution), equivalent to a reduction from $14293.863$~Mm$^2$ to $2266.325$~Mm$^2$. The second retained filament, in the data from 26 March 2016, shown in Figure~\ref{fig:ReissSamplesB}, was reduced from $591$~pixels to $51$~pixels ($512\times512$~pixel resolution), a reduction from $8110.057$~Mm$^2$ to $660.249$~Mm$^2$. QUACK03 resulted in the loss of 9 additional CH regions, 8 at the limb and a faint CH near disk center on 4 May 2019. Of these cases, 8 CHs were removed due to seed filtering, including the case near disk center. Removal of CHs at disk edge is not surprising given that the LOS magnetic field less accurately portrays unipolarity at disk edge. In the final case, on 2 October 2014, shown in Figure~\ref{fig:ReissSamplesC}, the CH on the NW edge of the limb is absent from the seed and was identified in ACWE03 due to evolution through the QS from an adjacent CH. Since QUACK03 prevented growth into the QS, this CH is absent in QUACK03. This CH is also absent in ACWE04 and QUACK04.

\begin{table}[t]
\centering
    \caption{Comparison of filaments present in ACWE04 and QUACK04 segmentations for the datset of \cite{reiss2023} showing the official label, number of pixels of each filament region at $512\times512$~pixel resolution, and total area in Mm$^2$.}
    \begin{tabular}{ccrrrr}
        \hline
        \textbf{Date} & \textbf{Label} & \multicolumn{2}{c}{\textbf{ACWE04 Size}} & \multicolumn{2}{c}{\textbf{QUACK04 Size}} \\ 
        \hline
        {} & {} & {Pixels} & {Mm$^2$} & {Pixels} & {Mm$^2$}\\
        {2014-07-15} & {Fil3} & {$523$} & {$7732.382$} & {$276$} & {$4049.278$}\\
        {2015-01-21} & {Fil1} & {$214$} & {$6874.345$} & {$17$} & {$542.313$}\\
        {2015-02-10} & {Fil1} & {$545$} & {$23922.795$} & {$95$} & {$6813.639$}\\
        {2015-02-10} & {Fil4} & {$316$} & {$17609.477$} & {$26$} & {$814.206$}\\
        {2015-02-10} & {Fil7} & {$1365$} & {$18021.584$} & {$671$} & {$8677.054$}\\
        {2015-03-31} & {Fil2} & {$472$} & {$8455.888$} & {$43$} & {$652.559$}\\
        {2015-04-18} & {Fil2} & {$185$} & {$6751.238$} & {$33$} & {$1400.311$}\\
        {2015-08-11} & {Fil2} & {$1486$} & {$18654.977$} & {$297$} & {$3767.503$}\\
        {2015-08-11} & {Fil3} & {$365$} & {$4972.121$} & {$48$} & {$641.910$}\\
        {2015-09-23} & {Fil1} & {$2212$} & {$31573.298$} & {$176$} & {$2184.171$}\\
        {2015-09-23} & {Fil2} & {$1712$} & {$25406.038$} & {$157$} & {$2255.122$}\\
        {2015-09-23} & {Fil3} & {$300$} & {$4437.861$} & {$36$} & {$469.649$}\\
        {2016-01-14} & {Fil2} & {$480$} & {$7811.111$} & {$27$} & {$411.945$}\\
        {2016-03-26} & {Fil1} & {$687$} & {$9751.211$} & {$383$} & {$5216.005$}\\
        {2016-03-26} & {Fil3} & {$197$} & {$3013.165$} & {$15$} & {$210.407$}\\
        {2016-12-28} & {Fil2} & {$46$} & {$723.455$} & {$14$} & {$200.016$}\\
        {2017-10-04} & {Fil2} & {$50$} & {$540.241$} & {$21$} & {$305.042$}\\
        {2017-10-04} & {Fil3} & {$83$} & {$1967.337$} & {$32$} & {$747.750$}\\
    \end{tabular}
    \label{tab:ACWEQUACK04}
\end{table}

QUACK04 eliminated 6 of the 24 filaments found in ACWE04. Effects on the remaining filaments are summarized in Table~\ref{tab:ACWEQUACK04}. We note that our tally of filaments for ACWE04 differs from those in \cite{reiss2023}  due to a difference in our count of filaments on 26 March 2016. Seeding with $\alpha=0.4$ appears to have significantly reduced the effectiveness of seed filtering. Exploring the relationship between $\alpha$ and the unipolarity threshold used in seed filtering should thus be an area of future research. Seed filtering also resulted in the loss of 5 CH regions, all at disk edge. We provide a summary of ACWE and QUACK performance in Table~\ref{tab:ReissMetrics}. All areas are first-order approximations assuming the area of each pixel $A_i$ of the full-scale image is
\begin{equation}
    A_i = \frac{A_{\rm{nom}}}{\cos(\theta_{i})},
    \label{eq:areaEst}
\end{equation}
where $A_{\rm{nom}}=0.189$Mm$^2$ is the nominal area of an AIA pixel at disk center \citep{lemen2012} and $\theta_{i}$ is the angular distance from the Earth-Sun line.

\begin{table}[t]
\centering
    \caption{Comparison of methods for \cite{reiss2023} dataset, outlining number of CHs correctly identified, number of CHs missed, number of filaments incorrectly segmented, number of filaments correctly ignored, and number of other object or features caught by each algorithm.}
    \begin{tabular}{cccccc}
        \hline
        \textbf{Method} & \multicolumn{2}{c}{\textbf{Coronal Holes}} & \multicolumn{2}{c}{\textbf{Filaments}} & \textbf{Other}\\
        \hline
        {} & {Identified} & {Missed} & {Identified} & {Ignored} & {Identified} \\
        {ACWE03} & {71} & {15} & {5} & {66} & {2}\\
        {QUACK03} & {63} & {23} & {2} & {69} & {2}\\
        {ACWE04} & {74} & {12} & {24} & {47} & {12}\\
        {QUACK04} & {69} & {18} & {18} & {53} & {10}\\
    \end{tabular}
    \label{tab:ReissMetrics}
\end{table}

\subsubsection{Extended Study of Segmentation Characteristics}
\label{sec:extended_study}

As additional validation, we use the dataset consisting of observations from the start of CR 2099 (13 July 2010) through the end of CR 2294 (2 March 2025), with a cadence of approximately 1 day between observations. We note, however, that a data gap of 8 days exists in CR 2180, a gap of 6 days in CR 2136, a gap of 4 days in CR 2207, and a gap of 2 days in CR 2243. In this extended dataset, segmentations are generated using the standard QUACK pipeline (Section~\ref{sec:pipelineSpecific}).

\begin{figure}
    \centering
    \includegraphics[trim=0in 0in 0in 0in,clip,width=0.93\textwidth]{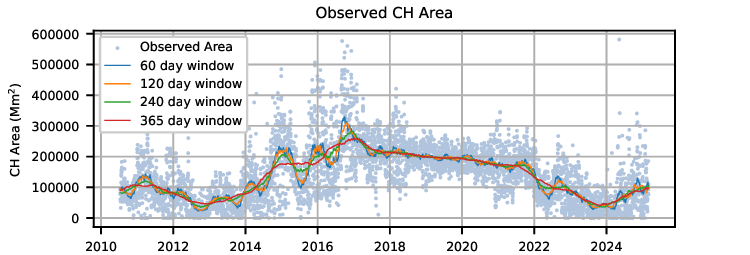}\\
    \caption{Total area of all regions identified as CHs using QUACK for each observation in the daily cadence dataset (dots) and mean area of all segmentations in a sliding window centered on the specified date (lines).}
    \label{fig:DailyCadence}
\end{figure}

Figure~\ref{fig:DailyCadence} summarizes the total area of each observation within the dataset, providing both the measured area of each segmentation in the form of a scatter plot and the mean area for sliding windows of various sizes (line plots). The measured area, provided in Mm$^2$, is the first-order approximation estimated using Equation~\ref{eq:areaEst}. We note that while the estimated CH area varies from one observation to the next, the means of the sliding windows indicate that CH area, as observed by our method, is lower during the two solar maxima and higher during the solar minimum. This is consistent with the expectation of fewer CHs during solar maximum. We further note that the current period of high solar activity is not present in either of the previous two datasets, and that no data after 22 February 2013 was used during the development of this algorithm, suggesting strong generalization of the QUACK pipeline.

For further study we divided each segmentation into the individual CH regions using the process outlined in Section~\ref{sec:SeedSplitAndOverfit}. This ensures that any disjoint regions along a CH boundary can be correctly associated with their respective CH. We calculated the area of each identified region and the location of its center of mass. We also calculated the mean magnetic field density $\bar{B}$ by first correcting for projection effects in the LOS observation $B$ on a pixel-by-pixel basis using
\begin{equation}
    B_{i\mathrm{Corr}} = \frac{B_{i}}{\cos(\theta_{i})},
    \label{eq:magEst}
\end{equation}
then calculating the total unbalanced magnetic flux of each region using
\begin{equation}
    \Phi=\sum_iB_{i\mathrm{Corr}}A_i,
    \label{eq:FluxTTl}
\end{equation}
and finally defining mean magnetic field density as
\begin{equation}
    \bar{B}=\frac{\Phi}{\sum_iA_i}.
    \label{eq:MagMean}
\end{equation}
For this, we utilized Level~1.5 HMI magnetograms, which have been aligned to the Level 1.5 AIA observations. This reprojection process was achieved by converting each magnetogram to a SunPy map object and using the \verb|reproject_to| method \citep{sunpy_community2020}. Alternatively, rescaling and aligning the segmentations to the Level~1 HMI magnetograms (and adjusting $A_{\rm{nom}}$ accordingly) produces highly similar results.

\begin{figure}
    \centering
    \includegraphics[trim=0in 0in 0in 0in,clip,width=0.93\textwidth]{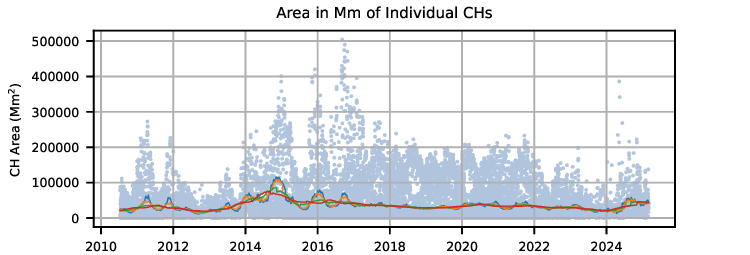}\\
    \caption{Total area of each region identified as a CH using QUACK for each observation in the daily cadence dataset (dots) and mean area of the individual regions in a sliding window centered on the specified date (lines). See Figure~\ref{fig:DailyCadence} for a legend.}
    \label{fig:DailyCadenceIndividual}
\end{figure}

\begin{figure}
    \centering
    \includegraphics[trim=0in 0in 0in 0in,clip,width=\textwidth]{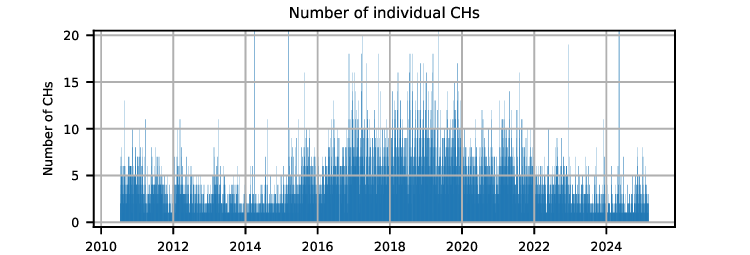}
    \caption{Number of CHs found in each observation. Note that six dates exceed the range of [0,20] of this plot: 30 March 2014, 12 March 2015, 7 May 2019, 5 May 2024, 08 May 2024, and 09 May 2024. All of these cases, except 7 May 2019, have exposure times $<0.15$s.}
    \label{fig:DailyCHs}
\end{figure}

\begin{figure}
    \centering
    \subfloat[All data within $\pm500^{\prime\prime}$]
    {\includegraphics[trim=0in 0in 0in 0in,clip,width=0.925\textwidth]{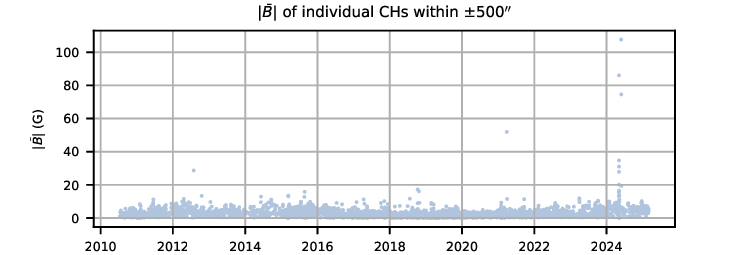}\label{fig:DailyCadenceMagneticA}}\\
    \subfloat[Zoomed in, with sliding window means]
    {\includegraphics[trim=0in 0in 0in 0in,clip,width=0.925\textwidth]{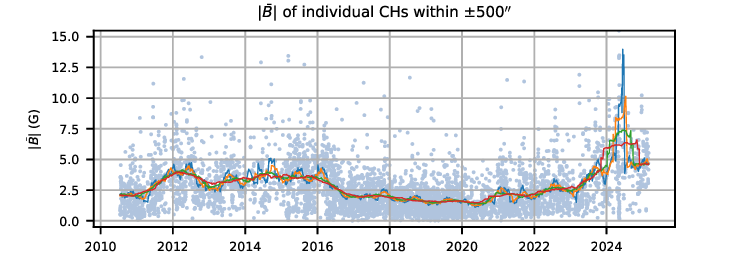}\label{fig:DailyCadenceMagneticB}}\\
    \caption{Absolute mean magnetic field density ($|\bar{B}|$) of individual CHs identified using QUACK for each CH within $\pm500^{\prime\prime}$ (dots) and in a sliding window centered on the specified date (lines). See Figure~\ref{fig:DailyCadence} for a legend of the line plots.}
    \label{fig:DailyCadenceMagnetic}
\end{figure}

In Figure~\ref{fig:DailyCadenceIndividual}, we provide a plot of the area in Mm$^2$ of all CH regions we identified.  Excluding local peaks caused by disk-edge CHs, we note that the mean area (for each sliding window) is relatively consistent across the dataset. This strongly indicates that the increase in CH area during solar minimum seen in Figure~\ref{fig:DailyCadence} is predominantly driven by an increase in the total number of CHs. This is confirmed in Figure~\ref{fig:DailyCHs}, which shows the total number of CH regions in each observation. Across the full dataset (including the cases at disk edge) we find that the mean area of individual CHs is $3.48\times10^{4}$~Mm$^2$ and the median area is $9.88\times10^{3}$~Mm$^2$.  This is similar to, but slightly higher than, the average area of $2.39\times10^{4}$~Mm$^2$ reported in~\cite{Hofmeister_2017}. 

Figure~\ref{fig:DailyCadenceMagnetic} shows the absolute mean magnetic field density ($|\bar{B}|$) of the regions with a center of mass within $\pm500^{\prime\prime}$ in the Helioprojective Cartesian system (dots) and means of sliding windows (lines). We note that the mean of $|\bar{B}|$, across all observations within the $\pm500^{\prime\prime}$ region is $2.60\pm3.04$~G.  Compared to the values of $2.9\pm1.9$G in~\cite{heinemann2019} and $2.97\pm1.55$G in~\cite{Hofmeister_2017}, we find a slightly lower average, indicating that QUACK is segmenting CHs with underlying magnetic field densities commensurate with expectations.  A potential source of difference may be that QUACK does not rely on an intensity threshold for CH delineation as in~\cite{Hofmeister_2017,heinemann2019}.  We further note a cyclical trend in $|\bar{B}|$ corresponding to the solar cycle (see Figure~\ref{fig:DailyCadenceMagneticB}), with slightly larger $|\bar{B}|$ around solar maximum and slightly lower $|\bar{B}|$ around solar minimum.  This is consistent with observations in~\cite{Hofmeister_2017} (and references therein).  The larger standard deviation found here as compared to~\cite{Hofmeister_2017,heinemann2019} is most likely due to the few outliers seen in Figure~\ref{fig:DailyCadenceMagneticA}.  These outliers are mainly concentrated in 2024 data, and correspond with significantly reduced exposure times, likely a result of solar flares associated with the Gannon storm. These outlier data are outside of the time range considered in~\cite{Hofmeister_2017,heinemann2019} so it is impossible to state whether those intensity-based algorithms would be similarly affected. Eliminating May 2024 from the computation results in a mean of $|\bar{B}|$ of $2.50\pm2.01$~G within $\pm500^{\prime\prime}$.

\section{Conclusions and Future Work}
\label{sec:conclusion}

Through the incorporation of magnetic field information, this new formulation of the ACWE algorithm, QUACK, is able to simultaneously consider region homogeneity in EUV and unipolarity in the magnetic field when delineating between CHs and other solar features. By relying on the fact that CHs appear as dark, homogeneous regions in EUV and as unipolar regions within the magnetic field, QUACK is able to produce more robust segmentations, retaining CH observations while simultaneously reducing the total area of false positive detections. In particular, QUACK reduced the presence of the surrounding QS (seen in CR 2099) and filaments (seen in CR 2133 and in the comparison to ACWE using the \cite{reiss2023} dataset).

Furthermore, we note that these benefits are retained even with aggressive spatial rescaling of both EUV and HMI magnetogram data to one-eighth of their original size (or $512\times512$~pixels). This rescaling offers two benefits. First, the ability to produce meaningful segmentations at a reduced spatial resolution greatly reduces computation time. Second, operating at reduced spatial resolution provides a reduction of filament detections since the thin structure of filaments increases the likelihood that dark pixels will not be retained at the lower resolution, eliminating them from the initial threshold-based seed outright. Even when not eliminated, however, the fact that meaningful segmentations are produced at a reduced spatial resolution ensures that filaments misidentified in the initial seed are still constrained, minimizing overall contamination. With the addition of seed filtering to eliminate non-unipolar regions, filament contamination can be further reduced without compromising identified CH regions.  This is evident in the results commensurate with \cite{Hofmeister_2017,heinemann2019} without the application of post-hoc processing to remove filament contamination as needed in those methods.

There are still several areas of potential improvement. First, the seed filtering process herein proposed can eliminate CHs near disk edge that would otherwise be correctly segmented. Addressing this issue may require a variable unipolarity threshold that takes into account the location of the region with respect to disk center. Second, this method relies on LOS magnetic field data, and we hope to explore the use of an estimate of radial magnetic field data to determine if this can further improve segmentations near the limb. Third, the broader relationship between initial seeding parameter and the ideal unipolarity cutoff for seed filtering should be explored to allow for reasonable unipolarity thresholds at values beyond standard seed of $\alpha=0.3$. Finally, we note that the previous formulation of ACWE~\citep{grajeda2023} provided the ability to quantify the uncertainty of detection based on the homogeneity of any given region compared to adjacent CH regions identified in the initial seeding process. We hope to reintroduce this ability in a future implementation QUACK, leveraging QUACK's ability to minimize inclusion of QS regions to minimize change of target and better characterize the CH region.
%
\section*{Acknowledgements}
The authors gratefully acknowledge the support of NASA grant 80NSSC20K0517, NASA Internal Scientist Funding Model (ISFM) awards to the Wang-Sheeley-Arge (WSA) model team, the Center for Helioanalytics (CfHA), the NSF Division of Atmospheric and Geospace Sciences grant AST-0946422 and multi-messenger astronomy supplemental funding to NSO, and NASA competed Heliophysics Internal Scientist Funding Model (ISFM), all of which helped support this work.

\section*{Data and Code Availability} 
The data used in this work are publicly available.  The GitHub repository at \url{https://github.com/DuckDuckPig/CH-ACWE} contains code to download the primary dataset used in this paper. The paper \cite{reiss2023} provides a link to the second dataset. The GitHub repository at \url{https://github.com/DuckDuckPig/CH-QUACK/tree/main/FinalPipeline/DatasetTools} contains code to download the final dataset used in this paper.  The GitHub repository at \url{https://github.com/DuckDuckPig/CH-QUACK} contains code to download the dataset used in this paper, the base ACWE and QUACK segmentation codes, and code to replicate all experiments described herein.

%
%
\bibliographystyle{plainnat}
\bibliography{References}  
%
%
%
%
\end{document}